%% file: becca_arxiv.tex
\documentclass{article}

\usepackage{arxiv}

\usepackage[utf8]{inputenc} 
\usepackage[T1]{fontenc}    
\usepackage{hyperref}       
\usepackage{url}            
\usepackage{booktabs}       
\usepackage{amsfonts}       
\usepackage{nicefrac}       
\usepackage{microtype}      
\usepackage{lipsum}		
\usepackage{graphicx}
\usepackage{natbib}
\usepackage{doi}
\usepackage{subcaption}
\usepackage{amsmath} 
\usepackage{amssymb}  
\usepackage{algorithm}
\usepackage{algpseudocode}
\usepackage{longtable} 
\usepackage[normalem]{ulem}

\usepackage{amstext,amssymb,amsfonts,latexsym}
\usepackage[utf8]{inputenc}
\usepackage[UKenglish]{babel}
\usepackage{csquotes}
\usepackage{ccicons}
\usepackage{animate}
\usepackage{setspace}
\usepackage{subcaption}
\usepackage{xcolor}
\usepackage{pdfpages}

\title{A {\bf Be}ta {\bf C}auchy-{\bf Ca}uchy (BECCA) shrinkage prior for Bayesian variable selection}

\author{ 
Linduni M. Rodrigo\textsuperscript{1,2},
Robert Kohn\textsuperscript{3},
Hadi M. Afshar\textsuperscript{1},
and Sally Cripps\textsuperscript{1,2}
\\
\\
\textsuperscript{1} Human Technology Institute, University of Technology Sydney, Ultimo NSW 2007.
\\
\textsuperscript{2}  Faculty of Engineering, University of Sydney, Camperdown NSW 2050.
\\
\textsuperscript{3}  University of New South Wales Business School, Kensington NSW 2033.
\\
}

\renewcommand{\shorttitle}{\textit{arXiv} Template}

\hypersetup{
pdftitle={A template for the arxiv style},
pdfsubject={q-bio.NC, q-bio.QM},
pdfauthor={David S.~Hippocampus, Elias D.~Striatum},
pdfkeywords={First keyword, Second keyword, More},
}

\newcommand{\V}[1]{\boldsymbol{\mathbf{#1}}}

\begin{document}

\maketitle

\begin{abstract}
	This paper introduces a novel Bayesian approach for variable selection in high-dimensional and potentially sparse regression settings. Our method replaces the indicator variables in the traditional spike and slab prior with  continuous, Beta-distributed random variables  and places  half Cauchy priors over the parameters of the Beta distribution, which significantly improves the predictive and inferential performance of the technique.  Similar to shrinkage methods, our continuous parameterization of the spike and slab prior enables us explore the posterior distributions of interest using fast gradient-based methods, such as Hamiltonian Monte Carlo (HMC), while at the same time explicitly allowing for variable selection in a principled framework. We study the frequentist properties of our model via simulation and show that our technique outperforms the latest Bayesian variable selection methods in both linear and logistic regression. The efficacy, applicability and performance of our approach,  are further underscored through its implementation on real datasets.

     \textbf{Keywords}: Bayesian methods, variable selection, shrinkage prior, linear regression, logistic regression, MCMC.
\end{abstract}


\section{Introduction}
\label{sec:Intro}
Identifying important predictors of an outcome, i.e. selecting variables, has been a central topic in statistical literature at least for 50 years \citep{Hurvich1989,Vach2001, Steyerberg2001, Meyer2019, Epprecht2021, Ullmann2024}, serving two key purposes: improving prediction accuracy and understanding or inferring the underlying structure of a system. In complex systems, where the number of possible covariates is  high, but where the actual number of covariates with nonzero effects on the response may be small, identifying relevant predictors becomes crucial \citep{Alhamzawi2018} to obtain a parsimonious yet flexible model.
Without variable selection, estimating models becomes computationally intensive,  and  predictions from these models may suffer from over-fitting, ultimately reducing the model's performance and interpretability \citep{Buhlmann2016, David2000}. 

In this paper we focus on Bayesian variable selection and compare our technique with other Bayesian methods. For a discussion on frequentist techniques, \citep{Fan2010,Huang2012,Jovic2015,Desboulets2018,Sauerbrei2020,Ullmann2024} give comprehensive  review.   In the Bayesian literature, variable selection techniques based on mixture distributions for priors on the regression coefficients, where a component of the mixture is a point mass at zero, has been an active area of research for over 40 years, \citep{zellner1986,Mitchell1988,Liang2008,Wang2015,Cao2021}. However, using a point mass as one of the components of the prior mixture distribution results in posteriors  with discontinuities. This in turn makes application of fast gradient-based sampling techniques such as Hamiltonian Monte Carlo (HMC) \citep{Duane1987} impossible, leaving the slower Markov chain Monte Carlo methods (MCMC) methods, such as Reversible Jump MCMC (RJMCMC) \cite{Green1995}, as the only viable alternative to estimate these posteriors. 

Bayesian shrinkage techniques overcome this computational burden by placing priors on regression coefficients that shrink values to zero rather than allow them to be identically zero. Examples of these techniques include the  Bayesian Lasso \citep{Park2008} which uses a Laplace distribution over the coefficients to shrink all coefficients, large or small, uniformly towards zero; the Horseshoe prior (HS) \citep{Carvalho2010} which allows the shrinkage to vary across the regression coefficients, with large coefficients to remain relatively unaffected, while shrinking smaller ones to zero; the Horseshoe+ (HS+) \citep{Bhadra2015} extends HS by adding an additional parameter, which allows for even further differential shrinkage; Dirichlet-Laplace (DL) prior \citep{Bhattacharya2015} which utilizes a Laplace distribution over the coefficients applying a joint shrinkage effect with the Dirichlet distribution.\\ 
\\
The continuous nature of these shrinkage priors, and hence continuous posteriors,  means that gradient-based methods can be used for fast exploration of, or approximation to, posteriors.  However variable selection is not explicitly encoded in the prior. Techniques such as the Lasso and DL, use posterior credible intervals with some user specified coverage to determine if the variable is zero or not \citep{Park2008,Zhang2018}. Additionally DL also assumes the existence of two clusters, those that are zero or non-zero, and uses {\it k-means} to identify these groups at each iterate in an MCMC chain \citep{Bhattacharya2015}. The HS and HS+ define a variable, $\kappa\in (0,1)$, which is the weight the posterior mean of the regression co-efficient places on zero and, again, this value can be used to classify regression coefficients as zero or non-zero.  However $\kappa$ is not the prior probability that a regression co-efficient is zero.  These priors focus more on the behavior of the coefficients rather than on explicit variable selection, which can make them less suitable for scenarios where model interpretability in terms of variable inclusion is critical. 
 For a full discussion on Bayesian variable selection see \citep{OHara2009,Lu2022},\\\\
Our contribution in this paper is threefold. First,  we propose a novel Bayesian approach for variable selection that yields better inferential and predictive results, by introducing a continuous Beta-distributed variable for the prior inclusion probability of a regression co-efficient - replacing the traditional binary indicator in spike-and-slab priors. Second, the method enables variable selection in a coherent framework, allowing robust inference for both individual predictors (marginal affect) and the combined influence of a group of predictors (joint effect). Third, it takes advantage of gradient based sampling techniques like Hamiltonian Monte Carlo (HMC), to ensure efficient computation and improved scalability in high-dimensional settings.\\
\\
The remainder of this paper is structured as follows: Section \ref{sec:Model&Prior} introduces the properties of the novel prior and outlines the model framework while Sections \ref{sec:simulation studies} and \ref{sec:Real applications} apply the proposed prior to linear and logistic regression, and show its performance in simulations and real-world applications.

\section{MODEL, PRIOR, \& POSTERIOR}
\label{sec:Model&Prior}
\subsection{Linear regression model}
Suppose we have an $n\times 1$ vector of responses, $\mathbf{y}=(y_1,\ldots, y_n)$, and corresponding observations on $p$ predictor variables in $\mathbf{X}=(\mathbf{x}_1,\ldots,\mathbf{x}_p)$, where $\mathbf{x}_j=(x_{1j},\ldots,x_{nj})'$ for $j=1,\ldots,p$. In linear regression, the relationship between  $\mathbf{y}$ and the predictors in $\mathbf{X}\in \mathbb{R}^{n \times p}$ is given by:
\begin{equation}
    \V{y} = \V{X}\V{\beta} + \V{\varepsilon} \label{eq:linear_model}  
\end{equation}
where $\V{\beta} = (\beta_1,\beta_2,...,\beta_p)^\top$ is the $p \times 1$ vector of regression coefficients and $\V {\varepsilon} \sim \mathcal{N}(\V{0}, \sigma^2\mathbf I_n)$ is a vector of independent and identically distributed error terms. We assume that all variables are zero-centered and exclude an intercept $\beta_0$ from the model given by \eqref{eq:linear_model}.

A common approach to Bayesian variable selection involves pairing each predictor variable $\mathbf{x}_j \in \mathbf{X}$ with a  binary latent variable, $I_j$, via a prior distribution, which is set to 1 or 0 depending on whether $\mathbf{x}_j$ is included or excluded in the model, for $j=1,\ldots,p$ \cite{Mitchell1988}. The model can then be represented as:
\begin{equation}
\mathbf{y} = \mathbf{X}_{A_1} \boldsymbol{\beta}_{A_1} + \boldsymbol{\varepsilon}
\end{equation}
where \( A_1 = \{ j ; \gamma_j =1,j = 1,2,...,p\}\), $p_1=||A_1||$, \( \mathbf{X}_{A_1} \) is the $n\times p_1 $ matrix of predictors that are included in the model, and 
 \( \boldsymbol{\beta}_{A_1} \) is the vector of regression coefficients corresponding to these predictors.
The likelihood is thus,
\begin{equation}
   f(\V y|\V \beta, \sigma^2,A_1)\propto \exp\left[-\frac{(\V y -  \V X_{A_1}\V \beta_{A_1})^\top (\V y -  \V X_{A_1}\V \beta_{A_1})}{2\sigma^2} \right]
   \label{eqn:linear_like}
\end{equation}
\subsection{Logistic regression model}
Consider the case where the response variable $\V y$  is binary i.e., $\V y = (y_1,...,y_n) \in \{0,1\}^n$ and follows a Bernoulli distribution with probability of success $\pi_i = \Pr(y_i=1); i= 1,...,n$. In logistic regression, the probability $\pi_i$ is given by,
\begin{equation}
    \pi_i=\frac{\exp(\V x_{iA_1}\V\beta_{A_1})}{1+\exp(\V x_{iA_1}\V\beta_{A_1})}
    \label{eqn:logit_prob}
\end{equation}
where $\V x_{iA_1}$ is the $i^{th}$ row of $X_{A_1}$, with the likelihood,
\begin{equation}
    f(\V y| \V X, \V \beta, A_1) = \prod_{i=1}^{n} f(y_i|\V X,\V \beta) = \prod_{i=1}^{n} \pi_i^{y_i}\left(1-\pi_i\right)^{1 - y_i}
    \label{eqn:logit_like}
\end{equation}

\subsection{Prior specification}
\label{subsec:Prior}
As discussed in Section~1, the SS prior for $\beta$ is a mixture of a continuous. and  discrete distribution. The discrete component makes exploring the posterior of $\beta$ challenging, for example by limiting the use of efficient gradient-based methods,  such as Hamiltonian Monte Carlo (HMC)\citep{Duane1987} and its variations \citep{Hoffman2011}.  \\
To justify our parameterization of the BECCA prior, we note that if the "slab" in the SS prior is a normal distribution with zero mean and variance of $c\sigma^2$, then $\beta\sim N(0,c\sigma^2)\gamma+\delta(0)(1-\gamma)$, where $\gamma=\Pr(\beta\ne 0)$, with $E(\beta^2|c,\sigma^2)=c\sigma^2\times\gamma=g\sigma^2\gamma^2$ if $c=g\gamma$.\\
The BECCA prior specified in (6), has a continuous distribution for $\beta$, however it is constructed so that the second moments of the SS and BECCA priors  are equal (as are the first and third, trivially). This is achieved by choosing the variance of $\beta$ in BECCA to be $g\sigma^2\gamma^2$. The BECCA prior specification is

\begin{eqnarray}
    \beta_{j} \mid \gamma_{j}, \sigma^2, g & \sim &\mathcal{N}(0,g\sigma^2\gamma_{j}^2) \label{eq:main_model} \nonumber \\
    \gamma_j \mid u,v &\sim& \text{Beta}(u, v) \nonumber\\
    g, u, v  &\sim &C^+(0,1)\nonumber\\
    \sigma^2&\sim& \text{IG}(a,b)
    \label{eqn:priors}
\end{eqnarray}
where $C^+(0,1)$ is the half-Cauchy prior.  The error variance $\sigma^2$ is assumed to have an inverse-gamma distribution with known hyper-parameters $a > 0$ and $b > 0$.  
As in the Horseshoe  \citep{Carvalho2010} and Horseshoe+ \citep{Bhadra2015} priors, the prior given by Equation \eqref{eqn:priors}, which we call the BECCA , provides both local and global shrinkage of the regression parameters - local shrinkage via $\gamma_j$ and global shrinkage via the parameter $g$ and the hyper-parameters $u$ and $v$.

The hyper-parameters $u$ and $v$ play a critical role in determining the departure of the BECCA prior from the SS prior. 
Consider the simplification  $u=v$, and $\sigma^2=1$ then as $u\rightarrow\infty$, $\gamma{\to}\sim\delta(0.5)$, that is $\gamma$ becomes a point mass at 0.5 and the  marginal of $\beta$ has a normal distribution, $\beta|g {\to}\sim N(0,0.25g)$. Conversely as $u{\to} 0$ then  $\gamma{\to}0.5\delta(0)+0.5\delta(1)$, that is $\gamma$ becomes an indicator variable $\gamma\in\{0,1\}$, with equal probability and therefore  the BECCA prior becomes the SS prior, that is $\beta|g{\to}0.5\delta(0)+0.5 N(0,g)$, where $\Pr(\beta\ne 0)=0.5$. BECCA thus couples the advantages of a SS prior for Bayesian model averaging and multiple hypothesis testing with the computational gains of working in the continuous space and a prior shrinkage profile which allows greater shrinkage for those regression coefficients close to zero while leaving large coefficients untouched.  These advantages are now discussed.
\subsection{Advantages of BECCA}
\begin{figure}[htb]
\centering
\begin{subfigure}[t]{0.47\textwidth}
        \centering
\includegraphics[width=1\linewidth]{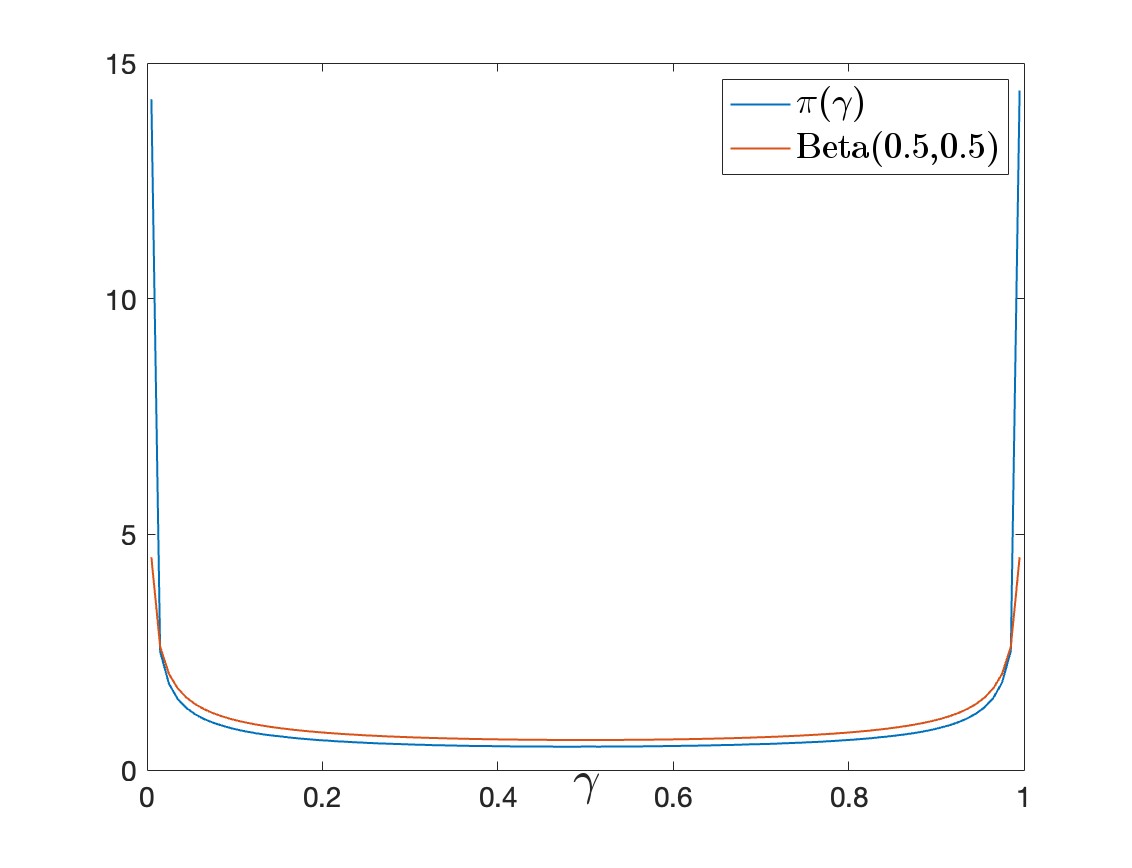}
\caption{}
\end{subfigure}
\begin{subfigure}[t]{0.47\textwidth}
        \centering
\includegraphics[width=1\linewidth]{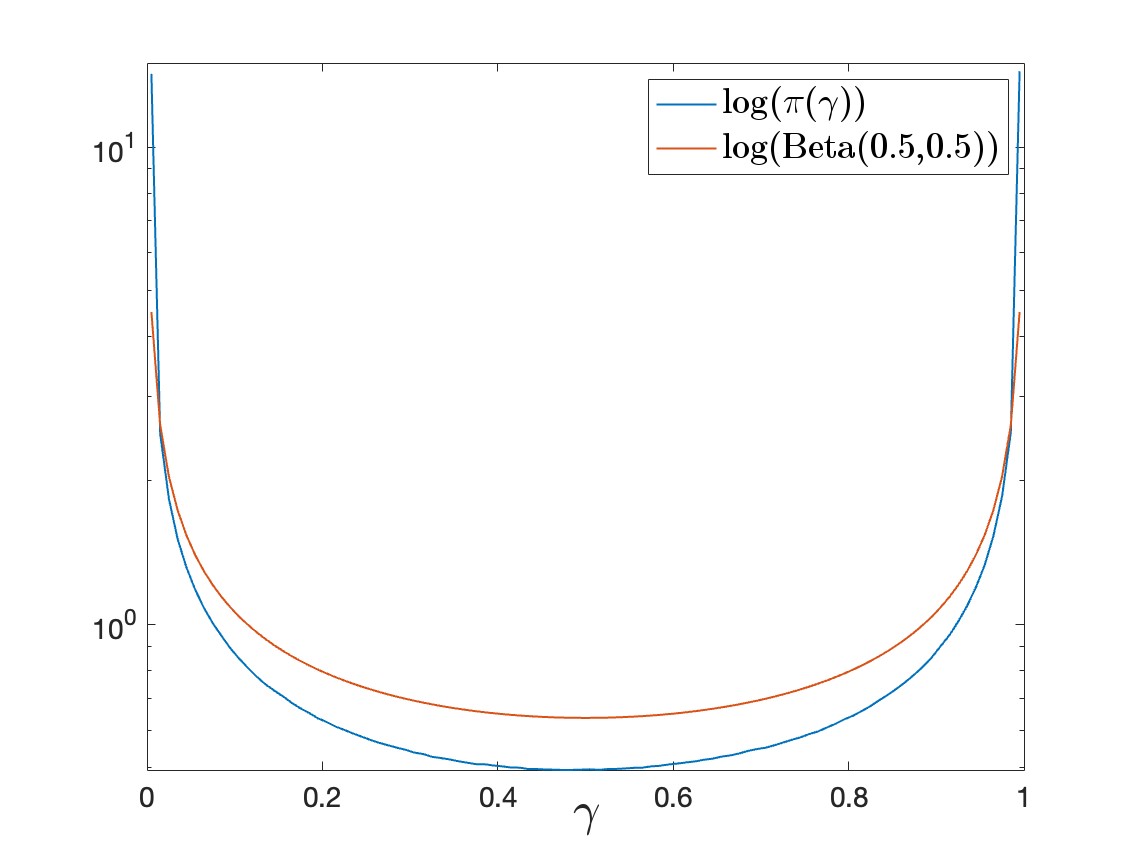}
\caption{}
\end{subfigure}
\caption{The marginal  prior density of $\gamma_j$ (blue) and a Beta(0.5,0.5) distribution (red). Panel~(a) is on a linear scale while panel (b) is on a log-scale.}
\label{fig:gamma_marg}
\end{figure}

{\bf (1). Shrinkage prior and profile.} The beta distribution on $\gamma_j$, with half Cauchy priors on the hyper-parameters $u$ and $v$ permits a flexible yet parsimonious model structure, and provides continuous posterior distributions of the quantities of interest.
The marginal density of \( \gamma_j \) is given by:
\begin{equation}
p(\gamma_j)=\int_0^{\infty}\int_0^{\infty} \frac{4\gamma_j^{u-1}(1 - \gamma_j)^{v-1}}{B(u, v)(1+u^2)(1+v^2)}dudv
   \label{eqn:gamm_prior}
\end{equation}
where \( B(u, v) \) is the Beta function.\\
Figure~\ref{fig:gamma_marg} panels~(a)~and~(b) compare this marginal distribution with the equivalent marginal distribution of the parameter $\kappa=\frac{1}{1+\lambda^2\tau^2}$ used in the HS, where $\kappa$ can be interpreted as a random shrinkage coefficient for the amount of weight that the posterior mean for $\beta$ places on zero once the data y have been observed \citep{Carvalho2010}.  \cite{Carvalho2010} assume that $\beta_j\sim N(0,\lambda_j^2\tau^2)$ with $\lambda_j\sim C^+(0,1)$, which implies that $\kappa_j|\tau^2=1\sim \mbox{Beta}(0.5,0.5$).  Figure~\ref{fig:gamma_marg}  shows that placing half-Cauchy priors over the hyper-parameters of a Beta distribution allows for additional shrinkage to either 0 or 1.  This increased flexibility results in improved shrinkage for covariates with coefficients close to zero, while leaving the larger coefficients untouched. 

\begin{figure*}[htb]  
\centering
\begin{subfigure}[b]{0.47\textwidth}  
    \centering
    \includegraphics[width=1\linewidth]{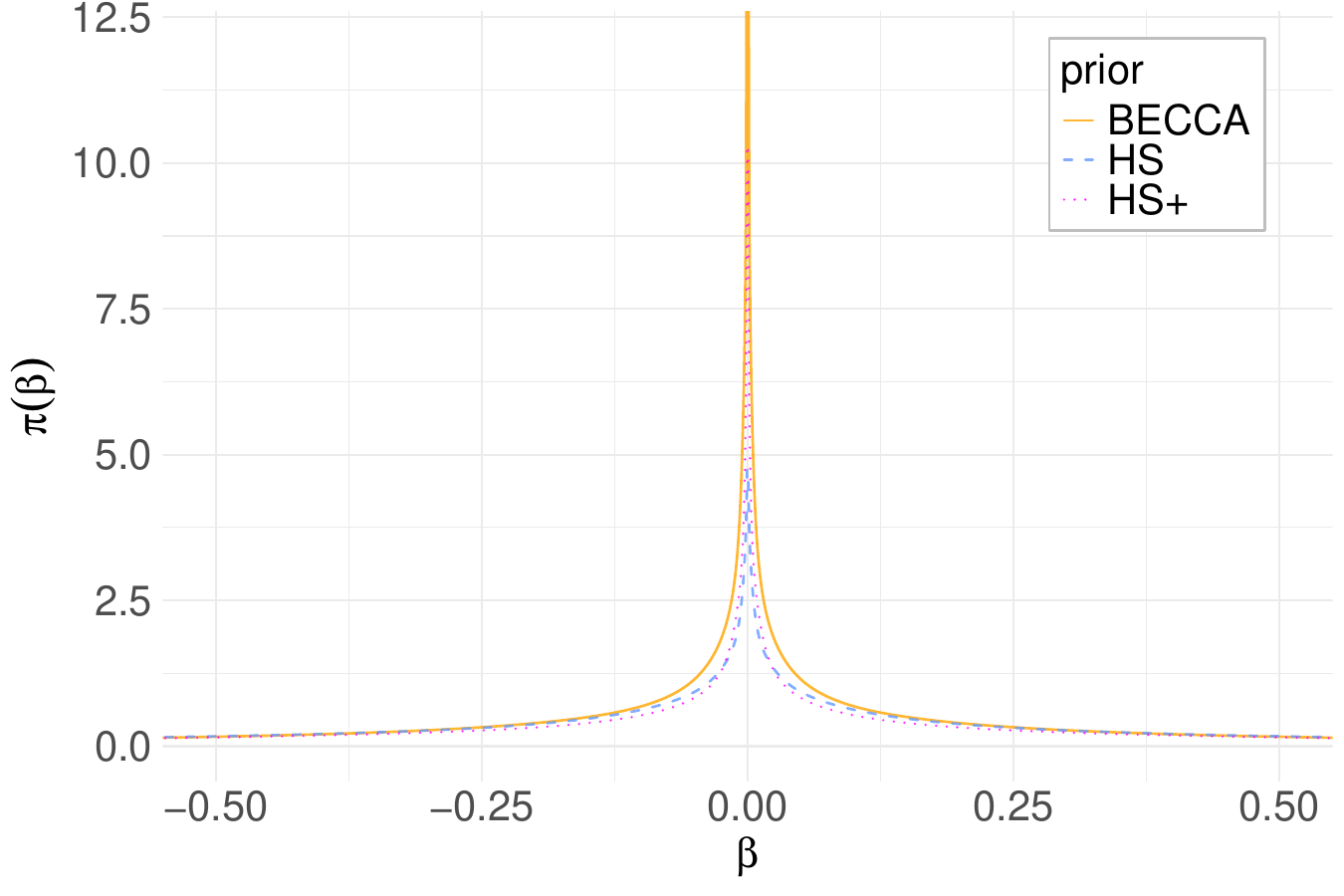}
    \caption{}
\end{subfigure}
\hspace{5mm}
\vspace{5mm}
\begin{subfigure}[b]{0.47\textwidth}  
    \centering
    \includegraphics[width=1\linewidth]{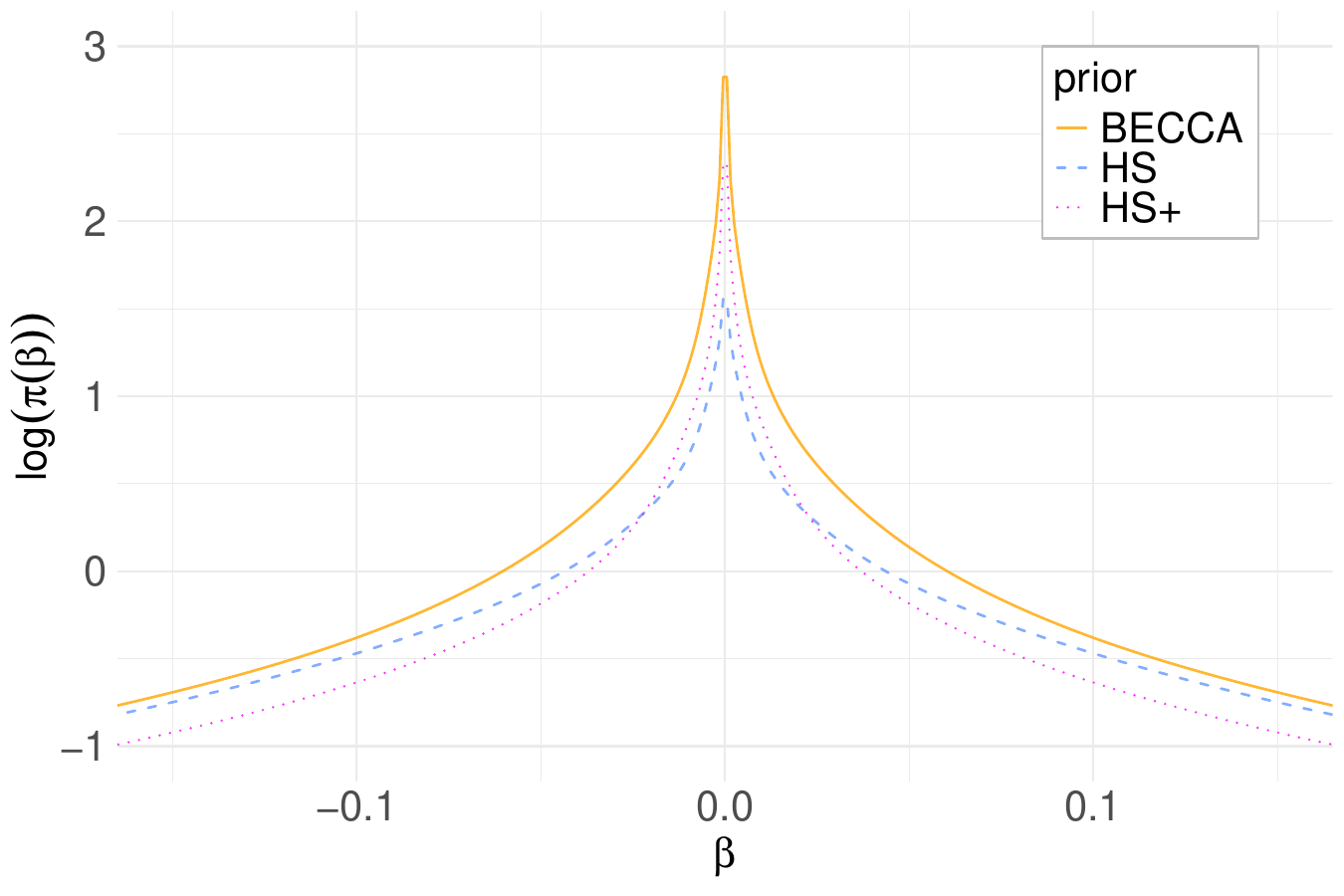}
    \caption{}
\end{subfigure}

\begin{subfigure}[b]{0.47\textwidth}  
    \centering
    \includegraphics[width=1\linewidth]{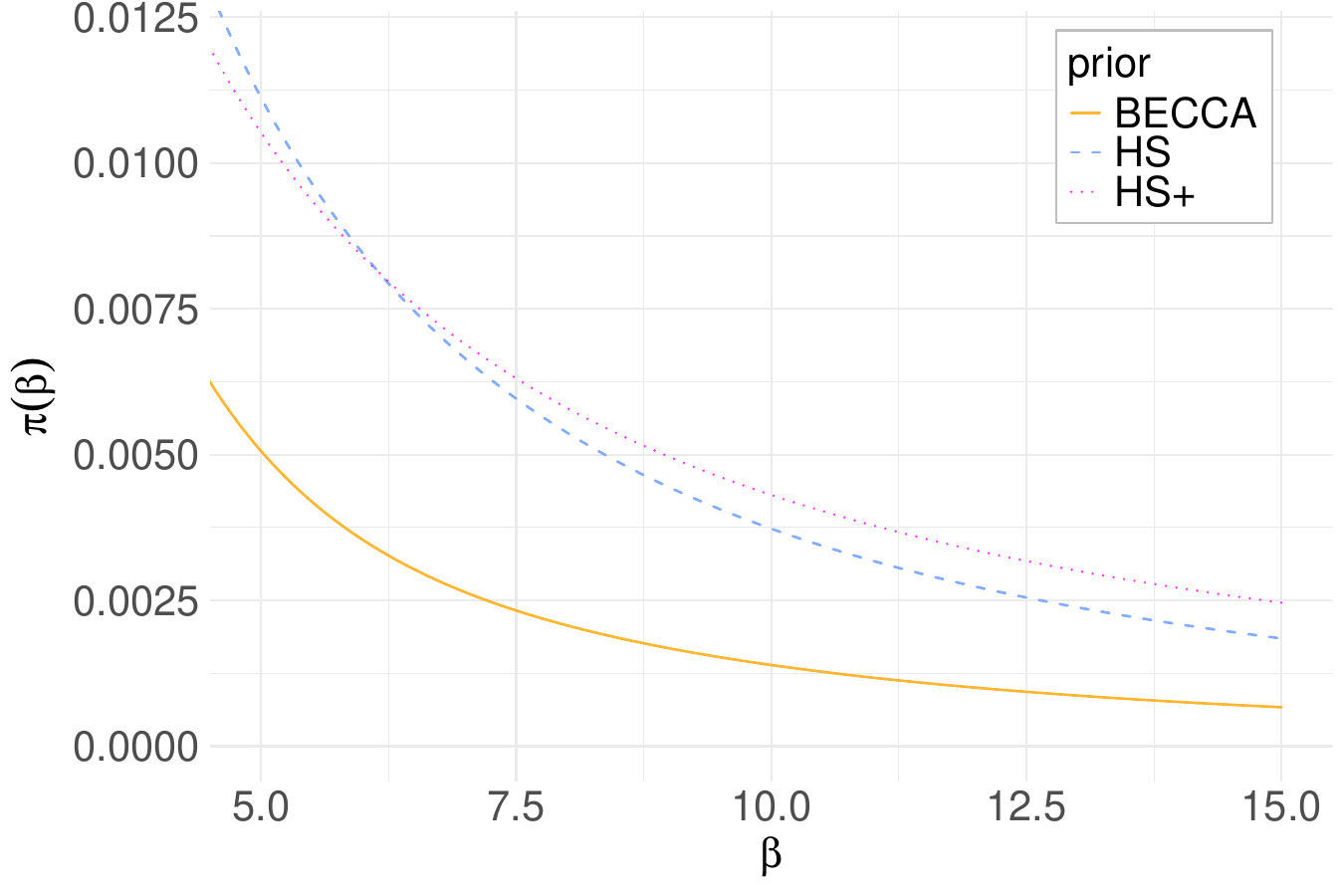}
    \caption{}
\end{subfigure}
\caption{Marginal prior densities of $\beta_j$ in \eqref{eqn:priors}. The legends denote the BECCA, HS, and HS+ priors. Panel (a) and (b) are on a linear and log scale respectively showing the region near the origin, while panel (c) is on a linear scale showing the tail region.}
\label{fig:beta_marg}
\end{figure*}

The marginal prior density, $\pi(\beta_j)$, corresponding to the priors in Equation \eqref{eqn:priors} is 
\begin{equation}
   p(\beta_j)=\int_{0}^{\infty}\int_0^{1} \frac{\exp\left\{-\frac{\beta_j^2}{2g\gamma_j^2}\right\}}{(2\pi g \gamma^2_j)^{1/2}(1+g^2)}p(\gamma_j)d\gamma_jdg
\end{equation}  
where $p(\gamma_j)$ is given by Equation \ref{eqn:gamm_prior}.  This does not have a closed-form expression, however, Figure~\ref{fig:beta_marg} panels (a)~-~(c) show plots of this marginal distribution for BECCA, HS, and HS + by numerically integrating over  $\gamma_j$ and $g$ for BECCA and over $\lambda_j$ and $\tau^2$ for the HS and HS+.

Figure~\ref{fig:beta_marg} panel (a)~-~(b) show that the BECCA prior shrinks values to zero more than the other priors (the BECCA prior has more weight around zero and its slope is steeper than that of the HS and HS+). Figure~\ref{fig:beta_marg} panel~(c), shows that the BECCA prior is almost constant for large values of $\beta_j$ indicating that the posterior $p(\beta_j|\mathbf y)\propto p(\mathbf y|\beta_j)p(\beta_j)$ is driven solely by the likelihood, and therefore large values of $\beta_j$, with likelihoods far from zero,  are less impacted by the BECCA prior than the HS and HS+ priors.
In short the BECCA prior provides more shrinkage for values of $\beta_j$ which are close to zero, than a either the HS or the HS+, while leaving the posterior unaffected by the prior for very large values of $\beta_j$.

\begin{figure*}[htb]
\centering
\begin{subfigure}[t]{0.31\textwidth}
        \centering
\includegraphics[width=1\linewidth]{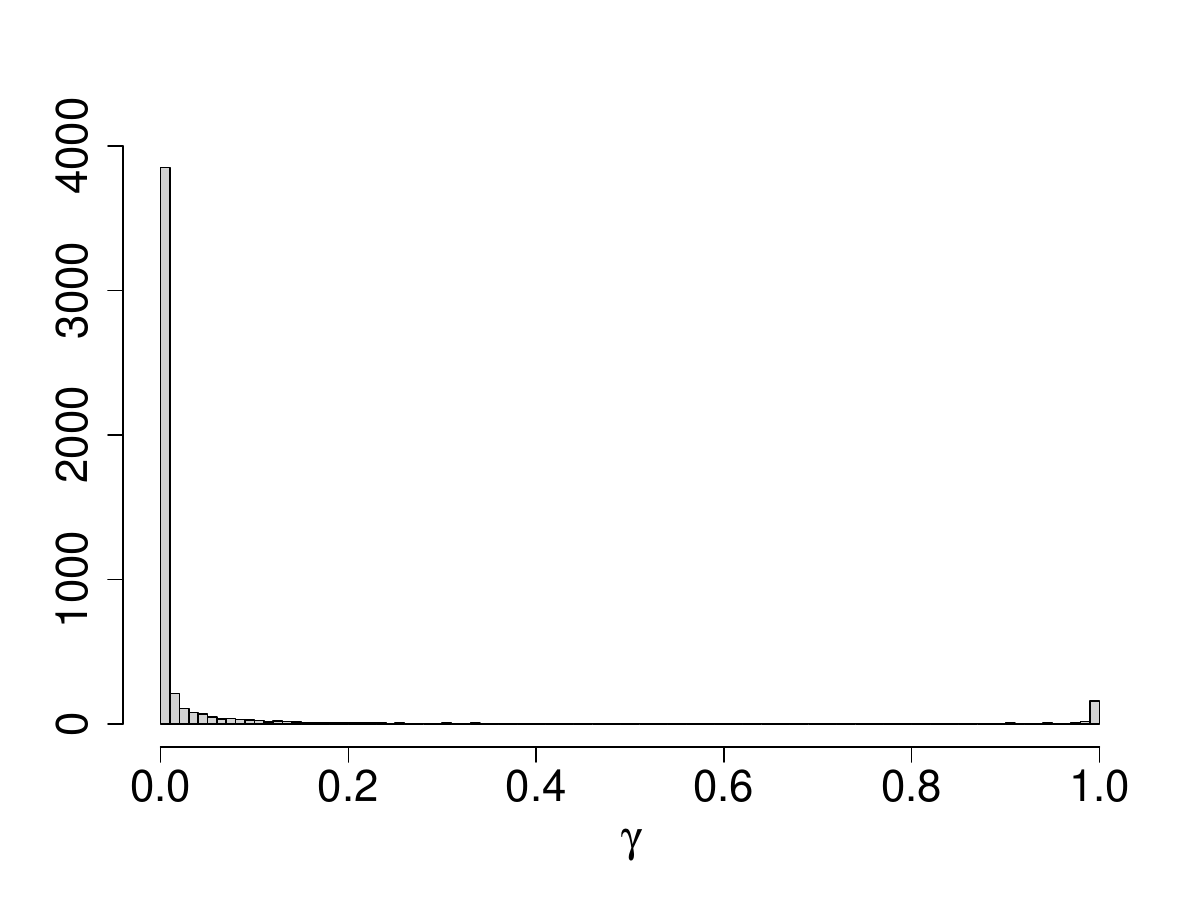}
\caption{}
\end{subfigure}
\begin{subfigure}[t]{0.31\textwidth}
        \centering
\includegraphics[width=1\linewidth]{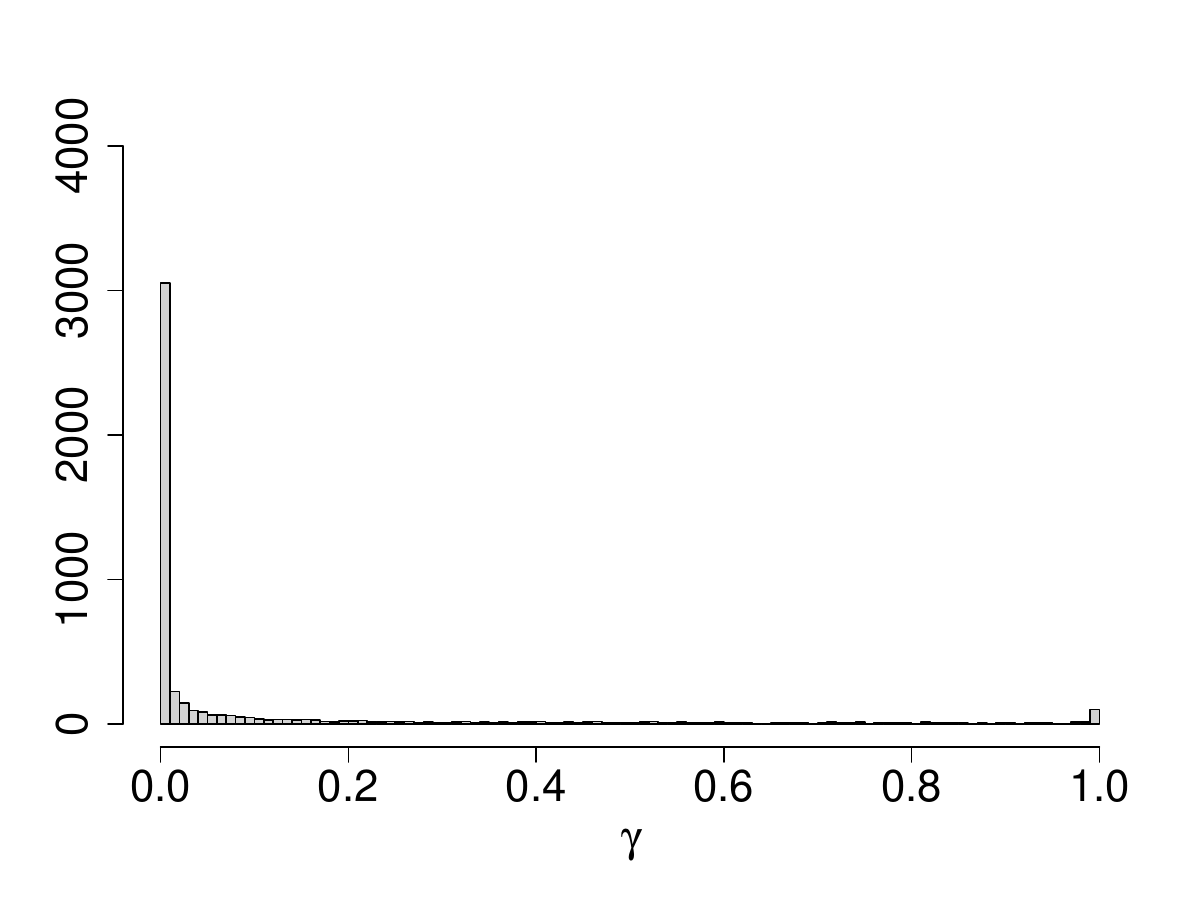}
\caption{}
\end{subfigure}
\begin{subfigure}[t]{0.31\textwidth}
        \centering
\includegraphics[width=1\linewidth]{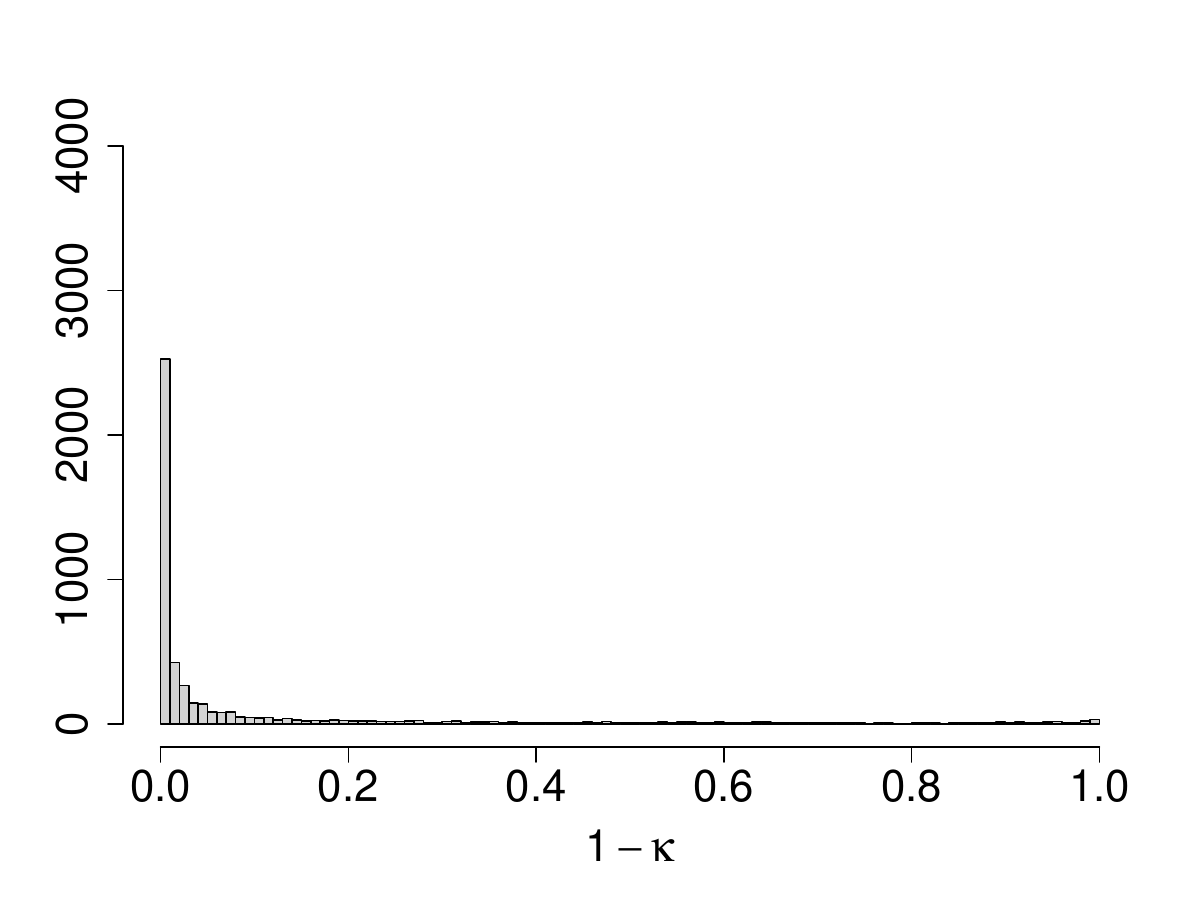}
\caption{}
\end{subfigure}
\caption{Posterior distribution of inclusion probability ($\gamma)$ for (a) BECCA and (b) SS with $1-\kappa$ for (c) HS in linear regression where the true $\beta$ is 0.}
\label{fig:Posterior_gamma_0}
\end{figure*}

\begin{figure*}[htb]
\centering
\begin{subfigure}[t]{0.31\textwidth}
        \centering
\includegraphics[width=1\linewidth]{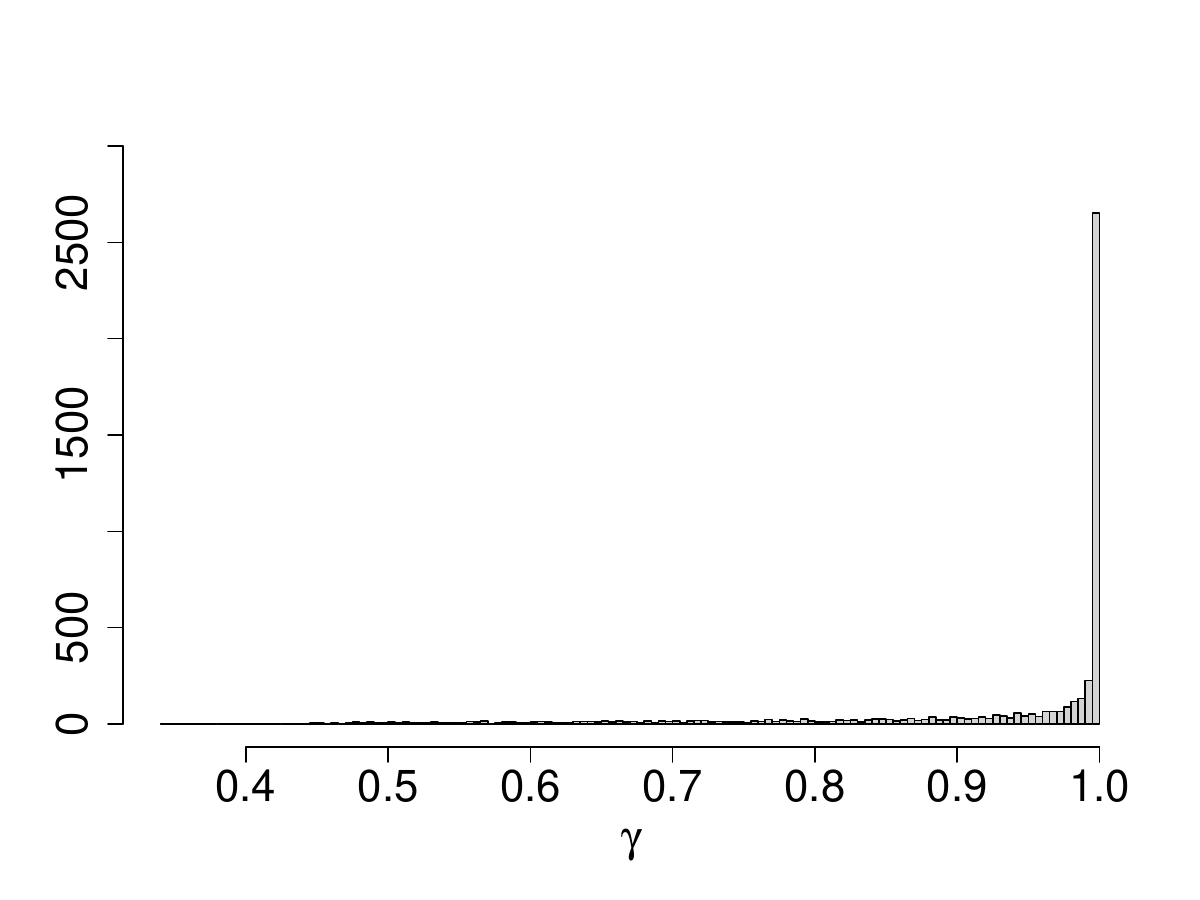}
\caption{}
\end{subfigure}
\begin{subfigure}[t]{0.31\textwidth}
        \centering
\includegraphics[width=1\linewidth]{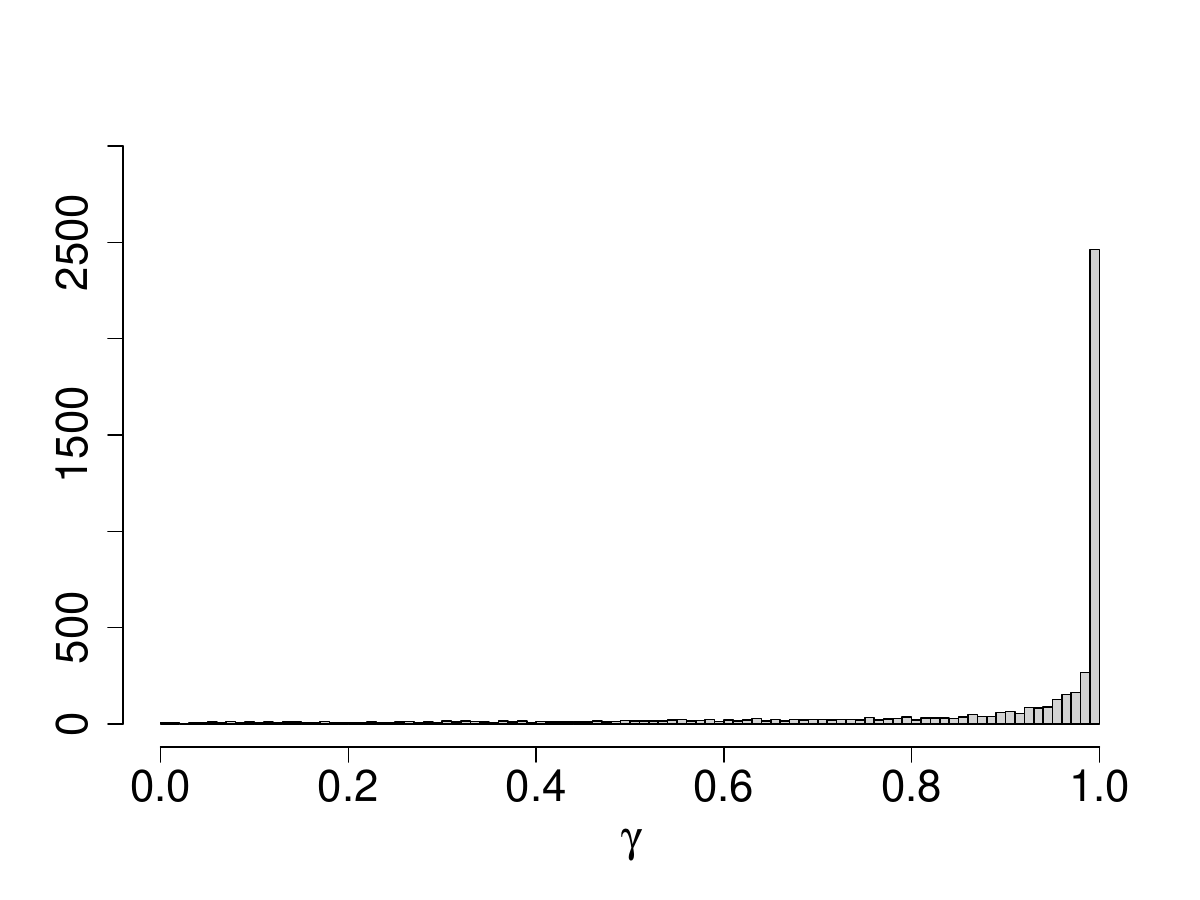}
\caption{}
\end{subfigure}
\begin{subfigure}[t]{0.31\textwidth}
        \centering
\includegraphics[width=1\linewidth]{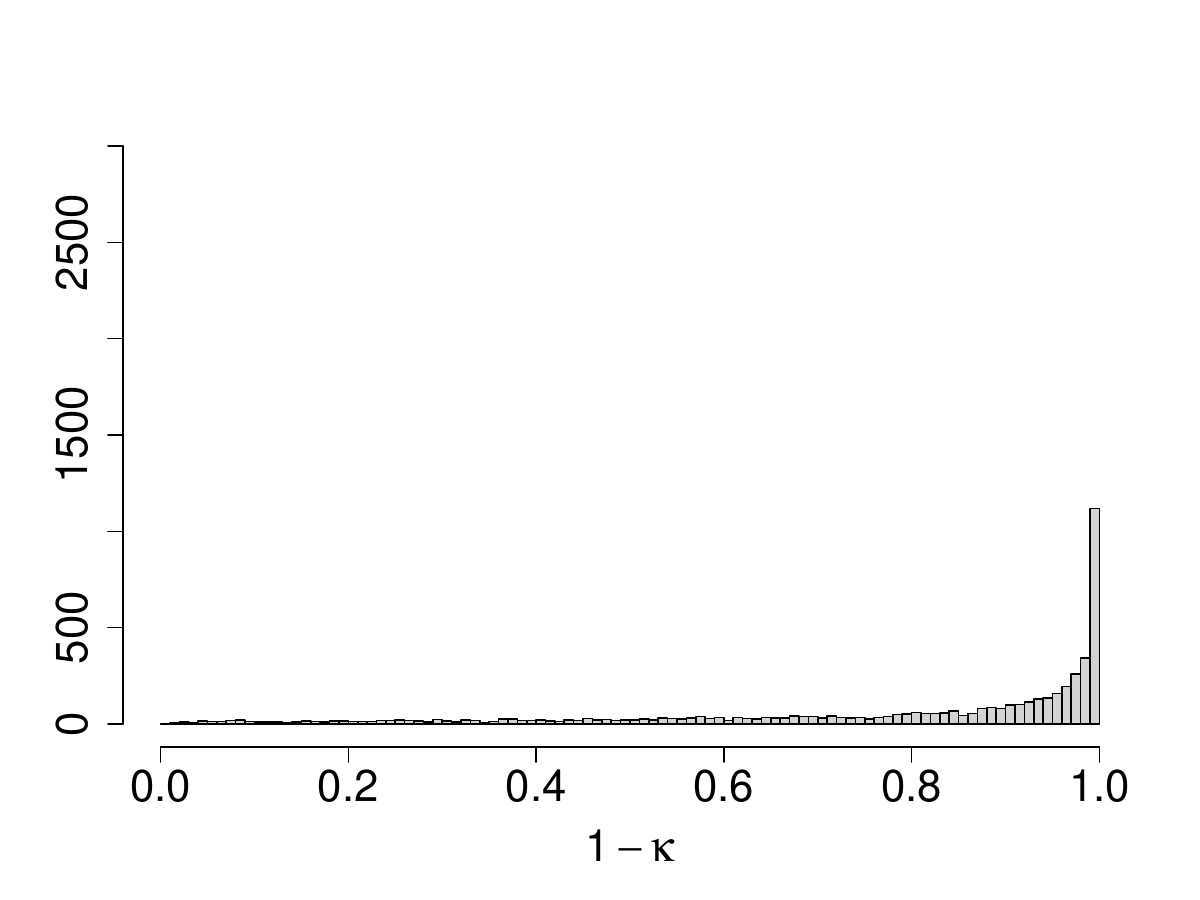}
\caption{}
\end{subfigure}
\caption{Posterior distribution of inclusion probability ($\gamma)$ for (a) BECCA and (b) SS with $1-\kappa$ for (c) HS in linear regression where the true $\beta$ is 2.5.}
\label{fig:Posterior_gamma_2.5}
\end{figure*}

We emphasis that  the motivation behind BECCA and the HS are different, with BECCA directly linking to the SS prior via the parameter $\gamma$, and that the BECCA prior can only be thought of as a HS prior with an additional hyper-parameter,  if $\tau^2=1$. In contrast the HS priors are placed on local ($\lambda$) and global ($\tau^2$) shrinkage parameters which induce a prior on the quantity $\kappa$, but as clearly stated in \cite{Carvalho2010}, $\kappa \ne 1-\Pr(\beta\ne 0)$. Figures~\ref{fig:Posterior_gamma_0}~and~\ref{fig:Posterior_gamma_2.5} show plots of the estimated posterior distribution of $\gamma_1$, and $\gamma_2$ and the HS equivalents, $1-\kappa_1$, $1-\kappa_2$, respectively, for the BECCA prior, panel~(a), the SS prior, panel~(b) and HS panel~(c) for a single realization of data generated from the model $y_i=x_{1i}\beta_1+x_{2i}\beta_2+e_i$, when $\beta_1=0,\beta_2=2.5$, and $(x_{1i},x_{2i})\sim N(0,I_2)$ and $i=1,\ldots n$, $n=100$.

{\bf (2). Correspondence with the SS Prior}. The second advantage of BECCA is that it provides a mechanism for automatically handling multiple hypothesis testing, (\cite{Scott2006}) and performing Bayesian model averaging.  We use HMC sampling to generate iterates from the posterior distribution of parameters given in Equation \eqref{eqn:priors}, namley $\beta$, $\sigma^2$, $g$ $\gamma$, $u$ and $v$, and by so doing obtain iterates of $\gamma=\Pr(\beta \neq0)$ in the SS formulation. We use the iterates of $\gamma$ to generate models consisting of subsets of variables, and thus are able to compute posterior probabilities of subsets of models. To see the equivalence between BECCA and SS prior, Figure~\ref{fig:Posterior_gamma_0} shows a plot of the estimated posterior distribution of $\gamma$, when $\beta=0$  for the BECCA prior, panel(a), the SS prior, panel(b). Figure~\ref{fig:Posterior_gamma_2.5} is an analogous plot when $\beta=2.5$

\subsection{Posteriors}
The joint posterior of the linear regression model corresponding to the priors in Equation \eqref{eqn:priors} is given by combining Equations~\ref{eqn:linear_like}~and~\ref{eqn:priors}. Similarly, the joint posterior of the logistic regression model is given by combining Equations~\ref{eqn:logit_like}~and~\ref{eqn:priors}.  Although  these  posteriors, given in Supplementary Materials, are analytically intractable, the derivatives of these posteriors w.r.t the parameters exist and therefore we use Hamiltonian Monte Carlo (HMC) \citep{Duane1987} with No-U-Turn Sampler (NUTS) \cite{Hoffman2011} implemented in \href{https://www.google.com/url?sa=t&rct=j&q=&esrc=s&source=web&cd=&cad=rja&uact=8&ved=2ahUKEwiS0Y2-xoyGAxXnrVYBHaZCCYYQFnoECBQQAQ&url=https%3A%2F%2Fmc-stan.org%2F&usg=AOvVaw1MjVekB5q9kJDP412xbWsM&opi=89978449}{Stan} to obtain a sampling estimate of the posterior.

\section{SIMULATION STUDIES}
\label{sec:simulation studies}
\label{subsec:simulation setting}
In this section, we evaluate and compare the performance of the BECCA prior with existing priors, the HS \citep{Carvalho2010}, HS+\citep{Bhadra2015} and DL \citep{Bhattacharya2015}, in terms of  inference - parameter estimation and variable selection, and prediction. We simulated data  for varying numbers of   predictors $p =$ 50, 100, and 200 and sparsity $q$ (i.e., the number of non-zero components).  For $p = 50$ and $100$, we set $q=10$ while for $p = 200$, we allow $q$ to be one of three values $q=10, 20, 30$. For each setting, we generate 50 datasets of size $n=100$. For details on the data simulation steps, HMC sampling procedure and diagnostics, please refer to the Supplementary Materials. 

{\bf Inferential performance:}
We evaluated the inferential performance of the priors in both linear and logistic regression using two metrics. The first is the Mean Squared Error (MSE) between the true coefficient value and the posterior sample mean $(\hat{\beta_j})$ given by;
\begin{equation}
    \text{MSE} = \frac{1}{p}\sum_{j=1}^p (\beta_j - \hat{\beta_j})^2
\end{equation}
The second metric is concerned with the proportion of times a method correctly identifies a regression coefficient as zero or non zero, and is measured by  sensitivity (also known as recall) and specificity \citep{Altman1994} and they are defined as,
\begin{align}
    \text{Sensitivity} &= \frac{\text{TP}}{\text{TP} + \text{FN}} \label{eq:sensitivity} \\
    \text{Specificity} &= \frac{\text{TN}}{\text{TN} + \text{FP}} \label{eq:specificity}
\end{align}
where TP, TF, FP and FN denotes True Positives, True Negatives, False Positive and False Negatives respectively.

For the BECCA prior we classified a coefficient as zero or nonzero based on the estimated posterior median  $\hat{\gamma}_j$, which is an estimate of $\Pr(\beta_j\ne 0|\V y)$.  If $\hat{\gamma}_j>0.5$ then we set $\beta_j \ne 0$, otherwise we set $\beta_j = 0$.  For the HS and HS+ \citep{Carvalho2010, Bhadra2015}, we take a similar approach and classify a coefficient as zero or nonzero based on the posterior median of  $\hat{\kappa}_j$ with the same classification rule. The DL prior \citep{Bhattacharya2015} was excluded from the variable selection comparison because it lacks a straightforward mechanism for assessing variable inclusion.

Tables \ref{tab:MSE_linear_beta} and \ref{tab:MSE_logit_beta} present the estimation accuracy in terms of the average MSE across 50 replicates along with the average difference between the performance of BECCA and other methods, and the standard errors of that difference, across various simulation settings. The lowest average MSE values among the different methods for each simulation setting are highlighted in bold. All simulation settings results for linear regression were obtained using a full equi-correlated design matrix, with correlation 0.75. Table \ref{tab:MSE_linear_beta} shows that in simulation settings with full equi-correlated design, BECCA's performance in estimating the true value of $\beta$ in a linear regression, as measured by the MSE, is significantly better than all other techniques.  Table~\ref{tab:MSE_logit_beta} gives the results for logistic regression and tells a similar story with one exception ($p=200$ and $q=20$),  BECCA's performance is not significantly better than DL. However it is interesting to note that, in linear regression setting, the superiority of BECCA over the HS decreases as the correlation between the covariates decreases.  For instance, in a simulation setting with $p=100$ and $q=10$ (starred in the Table \ref{tab:MSE_linear_beta}) with a sparse correlation matrix instead of full equi-correlated matrix, the HS (MSE = 0.904 ) estimation performance is significantly better than BECCA (MSE=0.907) with average difference 0.0032 and standard error of average difference 0.0003. This phenomenon can also be seen in the performance of the HS against the Spike and Slab Lasso \cite{Ročková2018}. 

Table \ref{tab-VariableSelectionLinear} presents the specificity and sensitivity  for linear regression. The results show that all priors perform similarly in terms of specificity indicating that all priors  identify almost all irrelevant variables. However, the sensitivity results show that the BECCA prior performs better and identifies non-zero coefficients in most data settings. It appears that the number of variables, or dimensionality, does not generally affect the performance of these variable selection methods. However, in the data setting III $(p=200,q=10)$, the sensitivity is relatively low across all methods. This could be attributed to the presence of many  coefficient with true values close to zero (i.e., $\beta_i = 0.028,0.421, -0.492$) and are consequently misidentified as noise. 

 \begin{table*}[htb]
\caption{Mean Squared Error (MSE) of posterior estimates of regression coefficients $\beta_j$ for BECCA, HS, HS+ and DL priors, in linear regression, computed over 50 replicates. Each data setting specifies the number of predictors ($p$) and the true number of non-zero predictors ($q$).  The improvement in performance provided by  BECCA over other techniques, measured as the MSE of the other technique minus the MSE of BECCA. Given in parentheses is average difference, first entry, and the standard error of the average difference, second entry. Average differences more than two standard deviations above zero are in bold. All simulation settings results were obtained using a full equi-correlated design matrix except for the starred one which was also evaluated using a sparse correlation matrix.} \label{tab:MSE_linear_beta}
\vspace{3mm}
\setlength\tabcolsep{15pt} 
\begin{center}
\begin{tabular}{@{}cccccc@{}} 
\toprule
 \textbf{p} & \textbf{q} &  \textbf{BECCA} &  \textbf{HS} &  \textbf{HS+} &  \textbf{DL}  \\ 
\midrule
50   &  10  & 0.060  & 0.183 ({\bf 0.123}, 0.006)  & 0.168 ({\bf 0.108}, 0.007) & 0.275 ({\bf 0.215}, 0.026)\\
\midrule
100*  &  10*  & $0.056$  & 0.172 ({\bf 0.116}, 0.005)  & 0.162 ({\bf0.106}, 0.005) & 0.061 ({\bf 0.005}, 0.002)\\
\midrule
200  &  10  & $0.167$  & 0.347 ({\bf 0.181}, 0.009) & 0.389 ({\bf0.222}, 0.012) & 0.894 ({\bf 0.727}, 0.136) \\
\cmidrule(lr){2-6}
200  &  20  & $0.192$  &  0.427 ({\bf 0.235}, 0.008)  & 0.491 ({\bf0.299}, 0.010) & 0.503 ({\bf0.311}, 0.012) \\
\cmidrule(lr){2-6}
200  &  30  & $0.567$  & 1.447 ({\bf 0.880}, 0.033)  & 1.539 ({\bf 0.972}, 0.037) & 2.167 ({\bf1.600}, 0.067) \\
\bottomrule
\end{tabular}
\end{center}
\end{table*}

The logistic regression results in table \ref{tab:VariableSelectionLogit} indicate that the BECCA prior better identifies irrelevant variables compared to HS and HS+ estimators, but that the HS and HS+ estimators perform better at identifying true relevant variables.

\begin{table*}[htb]
\centering
\caption{Mean Squared Error (MSE) of posterior estimates of regression coefficients $\beta_j$ for BECCA, HS, HS+ and DL priors, in logistic regression, computed over 50 replicates. Each data setting specifies the number of predictors ($p$) and the true number of non-zero predictors ($q$).  The improvement in performance provided by  BECCA over other techniques, measured as the MSE of the other technique minus the MSE of BECCA. Given in parentheses is average  difference, first entry, and the standard error of the average difference, second entry. Average differences more than two standard deviations above zero are in bold. }
\vspace{3mm}
\label{tab:MSE_logit_beta}
\setlength\tabcolsep{15pt} 
\begin{tabular}{@{}cccccc@{}} 
\toprule
 \textbf{p} & \textbf{q} &  \textbf{BECCA} &  \textbf{HS} &  \textbf{HS+} &  \textbf{DL}   \\ 
\midrule
50   &  10  & 0.650 & 0.783  ({\bf0.133}, 0.017) & 1.402 ({\bf0.751}, 0.128) & 0.691 ({\bf0.041}, 0.004) \\
\midrule
100   &  10  & 0.212 & 0.582 ({\bf0.371}, 0.041) & 1.479 ({\bf1.267}, 0.137) & 0.416 ({\bf0.204}, 0.015)\\
\midrule
200  &  10  & 0.499 & 1.378 ({\bf0.879},  0.379) & 1.897 ({\bf1.398}, 0.671) &  0.593 ({\bf0.094}, 0.010)\\
\cmidrule(lr){2-6}
200  &  20  & 0.458 & 0.829 ({\bf0.370}, 0.030) & 2.115 ({\bf 1.657}, 0.076) & 0.461 (0.003, 0.020) \\
\cmidrule(lr){2-6}
200  &  30   & 0.539 & 1.434 ({\bf0.895},  0.058) & 2.584 ({\bf2.045}, 0.116) & 0.614 ({\bf0.075},  0.015) \\
\bottomrule
\end{tabular}
\end{table*}

\begin{table}[htb]
\caption{Specificity (Spec) and Sensitivity (Sens), averaged across 50 replicates for BECCA, HS, HS+ and DL priors in linear regression, for varying number of predictors (p) and active predictors(q). The
highest values for each setting (in rows) are bolded and the empirical standard errors are reported in parentheses.}
\vspace{3mm}
\label{tab-VariableSelectionLinear}
\setlength\tabcolsep{10pt}
\begin{center}
\begin{tabular}{@{}cccccccc@{}} 
\toprule
& & \multicolumn{2}{c}{\textbf{BECCA}} & \multicolumn{2}{c}{\textbf{HS}} & \multicolumn{2}{c}{\textbf{HS+}} \\ 
\cmidrule(lr){3-4} \cmidrule(lr){5-6} \cmidrule(lr){7-8} 
\textbf{p} & \textbf{q} & \textbf{Spec} & \textbf{Sens} & \textbf{Spec} & \textbf{Sens} & \textbf{Spec} & \textbf{Sens} \\ 
\midrule
50  & 10 & 1 (0.000) & \textbf{0.898}(0.014) & 1 (0.000) & 0.596 (0.040)  & 0.999 (0.004) & 0.664 (0.078)\\
\midrule
100 & 10 & 1 (0.000) & \textbf{0.928} (0.011) & 1 (0.000) & 0.806 (0.024) & 1 (0.00) & 0.866 (0.066)\\
\midrule
200 & 10 & 1 (0.000) & 0.566 (0.052)         & 1 (0.000) & 0.600 (0.008) & 1 (0.000) & \textbf{0.614} (0.045) \\
\cmidrule(lr){2-8}
200 & 20 & 1 (0.000) & \textbf{0.950} (0.005) & 1 (0.000) & 0.742 (0.046) & 1 (0.000) & 0.773 (0.069)\\
\cmidrule(lr){2-8}
200 & 30 & 1 (0.000) & \textbf{0.859}(0.029) & 1 (0.000) & 0.766 (0.007) & 0.999 (0.001) & 0.773 (0.032)\\
\bottomrule
\end{tabular}
\end{center}
\end{table}

\begin{figure}[htb]
\centering
\begin{subfigure}[t]{0.46\linewidth}
    \centering
\includegraphics[width=0.95\linewidth]{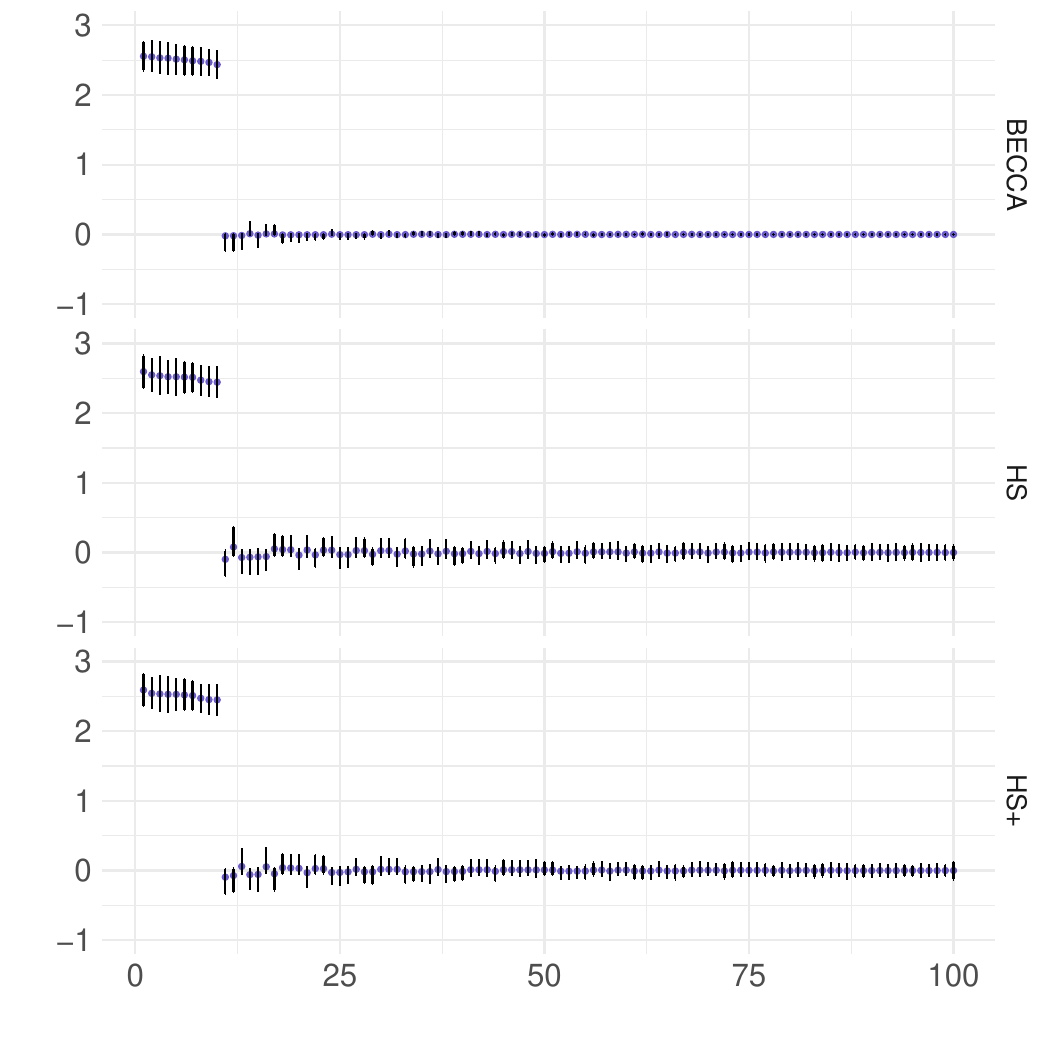}
    \caption{}
    \label{fig:95CI_allbeta}
\end{subfigure}
\hspace{1cm}
\begin{subfigure}[t]{0.46\linewidth}
    \centering
    \includegraphics[width=0.95\linewidth]{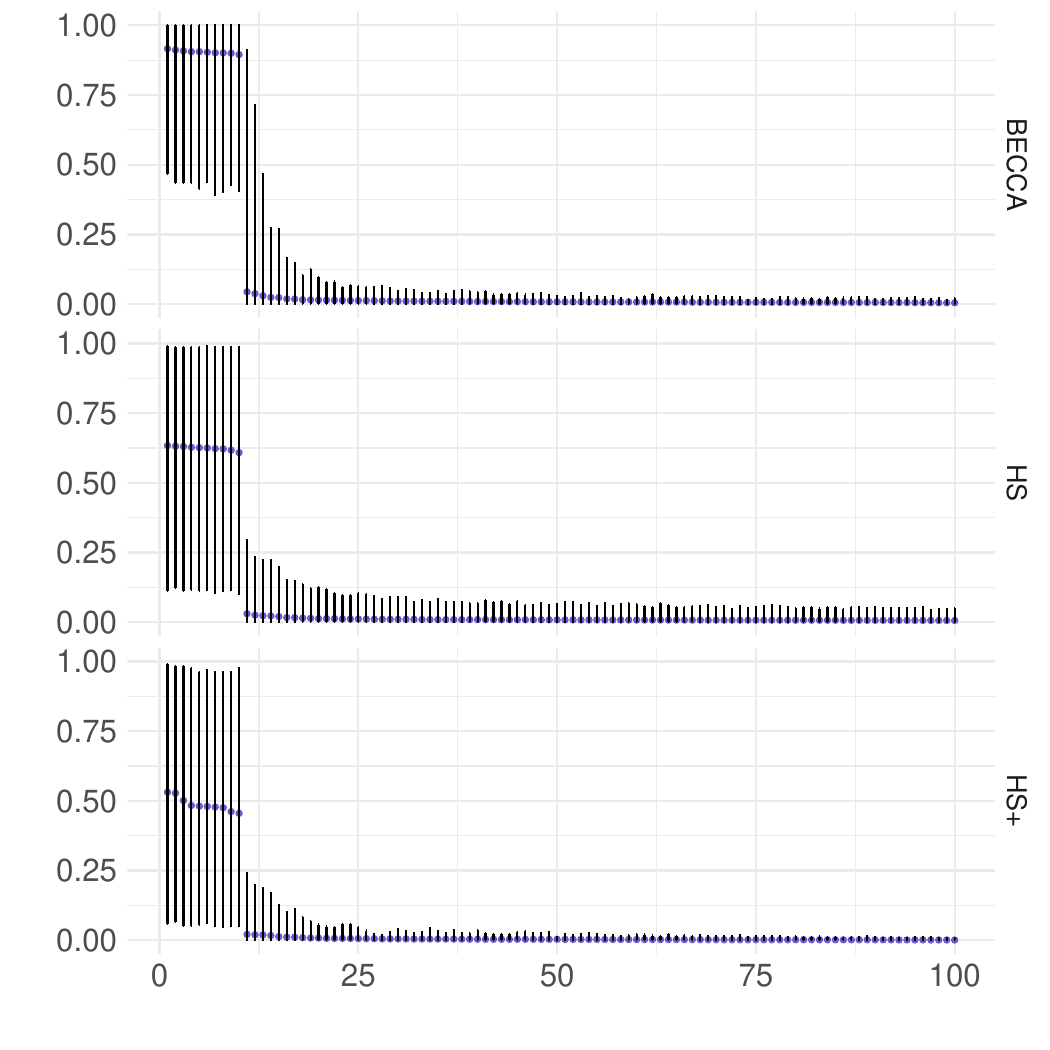}
    \caption{}
    \label{fig:95CI_gamma}
\end{subfigure}
\caption{Estimated (a) $\beta_j$ and (b) $\gamma_j$ for BECCA compared with estimated $1-\kappa_j$ for Horseshoe (HS) and Horseshoe+ (HS+) priors for 100 predictors with first 10 true $\beta_j$ equal to 2.5 with the rest equal to 0. Middle $95\%$ posterior credible intervals are depicted as black solid lines, and the posterior medians are shown in blue dots.}
\end{figure}

Figures \ref{fig:95CI_allbeta}~and~\ref{fig:95CI_gamma} present posterior estimates and corresponding 95$\%$ posterior credible intervals of linear model coefficients $\beta_j$ and inclusion $\gamma_j$ respectively from a single replicate of the data setting with $n=100$, $p=100$, and $q=10$.  All nonzero coefficients were set to 2.5  (see Supplementary Materials for additional comparisons, including results with the DL prior and scenarios with different true nonzero regression coefficient values). By reviewing the coefficient estimates across different methods, noting that the first 10 intervals correspond to intervals for true non-zero coefficients, followed by true zero coefficients, it is evident that, BECCA better estimates values closer to zero compared to both HS and HS+, while the estimates for true non-zero coefficients remain comparable across the methods. The estimates of the local shrinkage parameter indicate that BECCA outperforms HS methods in shrinking noise and identifying signals.

\begin{figure}[htb]
\centering
\begin{subfigure}[b]{0.3\textwidth}  
   \centering
   \includegraphics[width=1\linewidth]{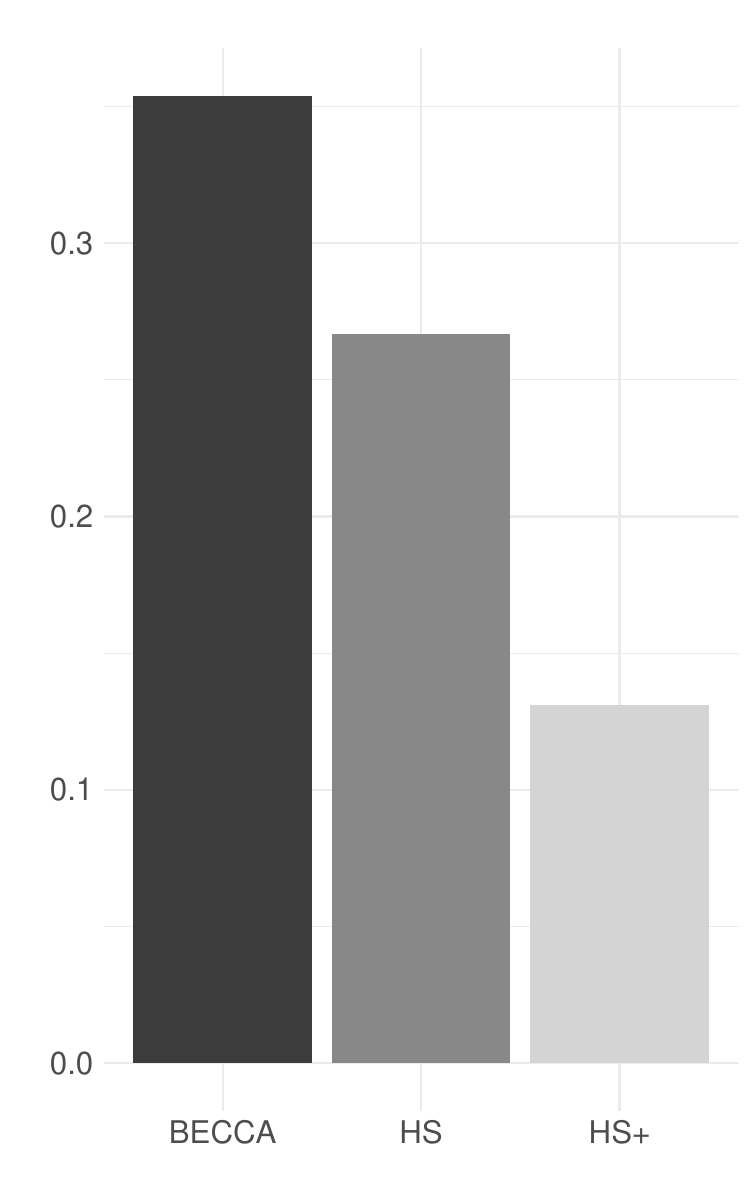 }
    \caption{}
\end{subfigure}
\hspace{1.5cm}
\begin{subfigure}[b]{0.5\textwidth}  
    \centering
    \includegraphics[width=1\linewidth]{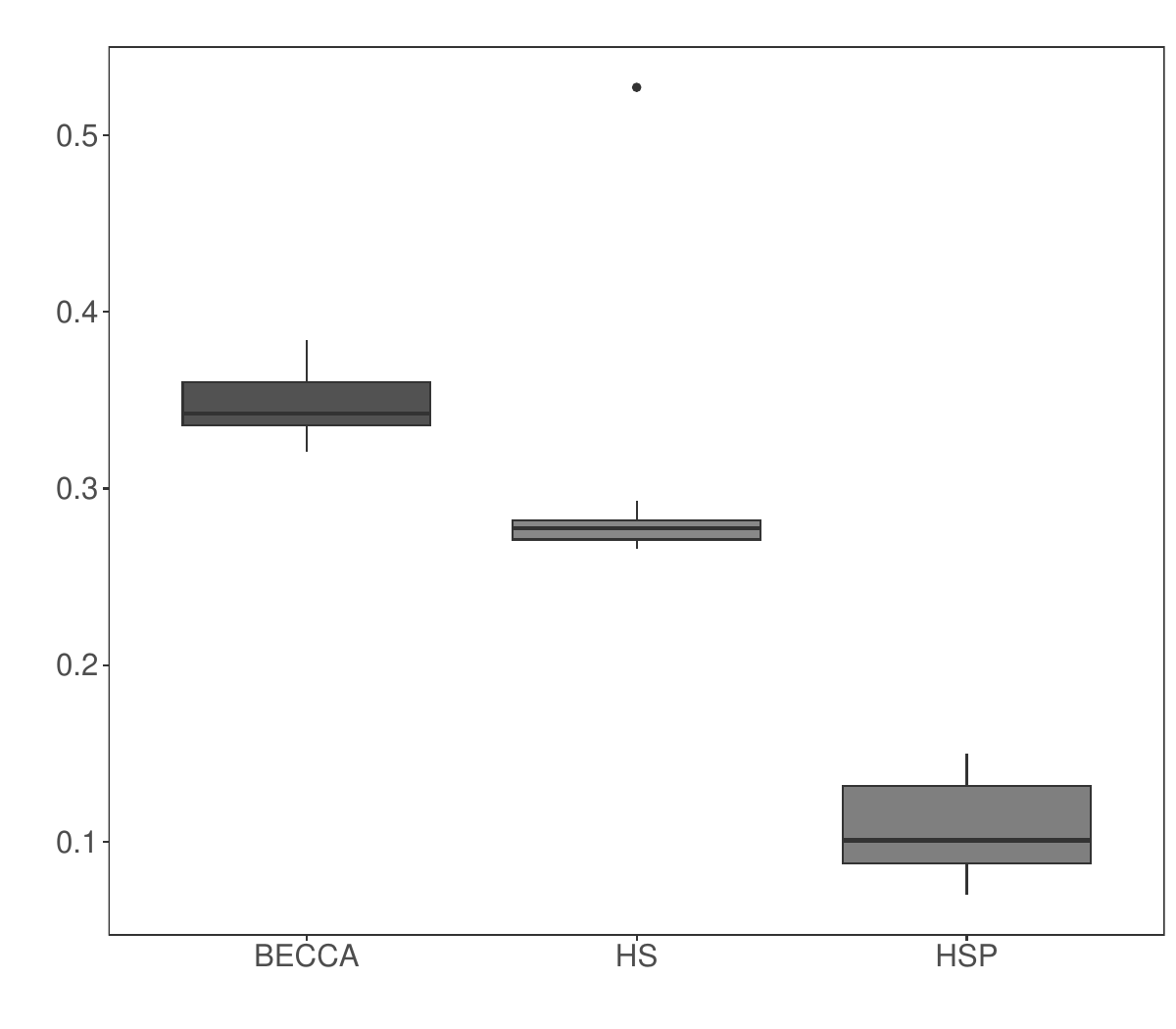}
    \caption{}
\end{subfigure}
\caption{\label{fig:Box_linear} Boxplots of  $\hat{\Pr}(\V I= \V I_{\mbox{true}})$ across 50 replications for the BECCA, HS and HS+ priors.}
\end{figure}

Figure~\ref{fig:Box_linear} shows the ability of each prior to 
identify the true set predictors.  To do this we define an 
indicator vector $\V I=(I_1,\ldots,I_p)$, where $I_j=1$ if 
$\beta_j\ne 0$ and $I_j=0$, otherwise, with $\V I_{\mbox{true}}$ being the true indicator vector from which the data were generated. At each iterate $k$ of the HMC, the iterate $\V I^{[k]}$ was generated using the values $\V\gamma^{[k]}$, noting that $\Pr(I_j=1)=\gamma_j$ for the BECCA prior, or using the values $\V\kappa^{[k]}$, noting that $\Pr(I_j=1)$ is analogous to $1-\kappa_j$ for the HS and HS+ priors. These iterates are used to compute an estimate  $\hat{\Pr}(\V I=\V I_{\mbox{true}})$ for each prior. Figure~\ref{fig:Box_linear}~panel(a) shows this quantity for a single realization, while panel~(b) shows a boxplot of this quantity for 50 realizations, for the three priors, and clearly demonstrates superior performance of BECCA over the HS and the HS+.
 

\begin{table}[htb]
\centering
\caption{Specificity (Spec) and Sensitivity (Sens), averaged across 50 replications for BECCA, HS, HS+ and DL priors in logistic regression, for varying
number of predictors ($p$) and active predictors($q$). The
highest metrics values for each setting (in
rows) are bolded and the empirical standard errors are reported in parentheses.}
\vspace{3mm}
\label{tab:VariableSelectionLogit}
\setlength\tabcolsep{8pt}
\begin{center}
\begin{tabular}{@{}cccccccc@{}} 
\toprule
& & \multicolumn{2}{c}{\textbf{BECCA}} & \multicolumn{2}{c}{\textbf{HS}} & \multicolumn{2}{c}{\textbf{HS+}} \\ 
\cmidrule(lr){3-4} \cmidrule(lr){5-6} \cmidrule(lr){7-8} 
\textbf{p} & \textbf{q} & \textbf{Spec} & \textbf{Sens} & \textbf{Spec} & \textbf{Sens} & \textbf{Spec} & \textbf{Sens} \\ 
\midrule
50   &  10  & \textbf{0.960} (0.024) & 0.700 (0.035) & 0.301 (0.054) & \textbf{0.968} (0.014) & 0.732 (0.257) & 0.910 (0.045)\\
\midrule
100   &  10  &  \textbf{0.999} (0.020) & 0.968 (0.058) & 0.566 (0.135) & 0.998 (0.018)  & 0.813 (0.082) & \textbf{1.000} (0.000) \\
\midrule
200  & 10 & \textbf{1.000} (0.000) & 0.865 (0.076) & 0.456 (0.347) & \textbf{1.000} (0.000) & 0.792 (0.286) & 0.953 (0.063) \\
\cmidrule(lr){2-8}
200 & 20 & \textbf{0.999} (0.005) & 0.860 (0.034) & 0.655 (0.312) & \textbf{0.913} (0.095) & 0.724 (0.342) & 0.839 (0.113)  \\
\cmidrule(lr){2-8}
200 & 30 & \textbf{0.993} (0.010) & 0.523 (0.032) & 0.447 (0.351) & \textbf{0.823} (0.161) & 0.728 (0.345) & 0.631 (0.185) \\
\bottomrule
\end{tabular}
\end{center}
\end{table}

{\bf Prediction performance:}
\label{subsec:Prediction}
To evaluate the predictive performance of the BECCA prior relative to the HS, HS+ and DL priors for both linear and logistic regression models we used the Mean Squared Prediction Error (MSPE) given by Equation \ref{eq:MSPE}. For each simulation setting described in \ref{subsec:simulation setting}, the dataset was partitioned into training and test set and we performed 5-fold cross validation. For each model, regression coefficients were estimated using the training set by averaging the posterior samples from different prior settings. The discrepancy between the estimated and observed outcome was calculated for each observation in the test set.
 We identify the covariate values in the test set by $\mathbf x_i^t.$. The MSPE given by,\\
\begin{equation}
    \text{MSPE}= \frac{1}{n^t}\sum_{i=1}^{n^t} (\mathbf{x}_{i.}^t\boldsymbol{\beta} - \mathbf{x}_{i.}^t\hat{\boldsymbol{\beta}})^2
    \label{eq:MSPE}
\end{equation}

where $n^t$ is the number of observations in the test set. For logistic regression, MSPE was computed as the deviation of estimated success probability, $\hat{\pi}_i^t$ from the true probability, $\pi_i^t$ where $\pi_i$ is defined in Equation \ref{eqn:logit_prob}. The MSPE values, averaged across 50 realizations, along with the average difference in prediction performance between BECCA and other methods, and the corresponding standard errors, are presented in Table \ref{tab:MSPE-linear-beta} for linear regression and Table \ref{tab:MSPE-logit-beta} for logistic regression. An MSPE value close to zero indicates strong prediction performance of a model.
\begin{table*}[htb]
\centering
\caption{Mean Squared Prediction Error (MSPE), averaged across 50 replicates in linear regression. Each data setting specifies the number of predictors ($p$) and the true number of non-zero predictors ($q$).  The improvement in prediction performance provided by BECCA over other techniques, measured as the MSPE of the other technique minus the MSPE of BECCA.  Given in parentheses is average difference, first entry, and the standard error of the average difference,
second entry. Average differences more than two standard deviations above zero are in bold. }
\vspace{3mm}
\label{tab:MSPE-linear-beta}
\setlength\tabcolsep{12pt} 
\begin{tabular}{@{}cccccc@{}} 
\toprule
 \textbf{p} & \textbf{q} &  \textbf{BECCA} &  \textbf{HS} &  \textbf{HS+} &  \textbf{DL}  \\ 
\midrule
50  & 10 & 0.0395  &  0.0819 ({\bf0.042},  0.003) & 0.0709 ({\bf0.031}, 0.003)  & 0.0688 ({\bf0.029}, 
 0.002) \\
\midrule
100 & 10 & 0.0239 & 0.0651 ({\bf0.045}, 0.004)  & 0.05271 ({\bf0.029}, 0.002) & 0.0733 ({\bf0.051}, 0.004) \\
\midrule
200 & 10 & 0.0940 &  0.1081 ({\bf0.015}, 0.003)  & 0.1167({\bf0.025}, 0.004)  & 0.2072 ({\bf0.116}, 0.009) \\ 
\midrule
200 & 20 & 0.09288 &  0.1365 ({\bf0.046}, 0.005)  & 0.1441 ({\bf0.051}, 0.005)  & 0.2451 ({\bf0.153}, 0.010) \\
\midrule
200 & 30 & 0.2639 &  0.7226 ({\bf0.449}, 0.028)  & 0.7267 ({\bf0.453}, 0.029)  & 1.5131 ({\bf1.243}, 0.121) \\
\bottomrule
\end{tabular}
\end{table*}

\begin{table*}[htb]
\centering
\caption{Mean Squared Prediction Error (MSPE), averaged across 50 replicates in logistic regression. Each data setting specifies the number of predictors ($p$) and the true number of non-zero predictors ($q$).  The improvement in performance provided by  BECCA over other techniques, measured as the MSPE of the other technique minus the MSPE of BECCA. Given in parentheses is average difference, first entry, and the standard error of the average difference,
second entry. Average differences more than two standard deviations above zero are in bold. }
\vspace{3mm}
\label{tab:MSPE-logit-beta}
\setlength\tabcolsep{14pt} 
\begin{tabular}{@{}cccccc@{}} 
\toprule
 \textbf{p} & \textbf{q} &  \textbf{BECCA} &  \textbf{HS} &  \textbf{HS+} &  \textbf{DL}   \\ 
 \midrule
50  & 10 & 0.036   & 0.049 ({\bf0.012}, 0.002)  & 0.042 ({\bf0.005}, 0.002)& 0.032 (-0.005, 0.002) \\
\midrule 
100  & 10 & 0.058 &  0.087 ({\bf0.029}, 0.004) & 0.071 ({\bf0.014}, 0.003)  & 0.070 ({\bf0.011}, 0.003) \\
\midrule
200 & 10 & 0.064  & 0.072 ({\bf0.008}, 0.003)  & 0.074 ({\bf0.010}, 0.003)  &  0.069 (0.0005, 0.003) \\
\midrule
200 & 20 & 0.133 & 0.164 ({\bf0.034}, 0.006)  & 0.145 ({\bf0.015}, 0.005)  & 0.135 (0.002, 0.003)  \\
\midrule
200 & 30 & 0.113  & 0.126 ({\bf0.013}, 0.004)  & 0.134 ({\bf0.021}, 0.004)  & 0.109 (-0.003, 0.003) \\
\bottomrule
\end{tabular}
\end{table*}

Table $\ref{tab:MSPE-linear-beta}$ shows that, regardless of the dimensionality $p$ and number of predictors with true non-zero coefficients $q$, BECCA performs significantly better than all other techniques (substantially so), achieving lowest MSPE. Table \ref{tab:MSPE-logit-beta} gives
the results for logistic regression. For logistic regression, BECCA’s performance is significantly better than HS
and HS+, however BECCA’s performance significantly exceeds that of DL in only one of the settings. 

In summary, the proposed BECCA prior performs comparably or better than other methods in coefficient estimation, variable selection and prediction for linear regression under varying data configurations and sparsity levels. In logistic regression, while the proposed method excels in estimation, identifying irrelevant variables and prediction, it is less effective at identifying signals compared to HS estimators.

\section{REAL DATA APPLICATIONS}
\label{sec:Real applications}
In this section we evaluate the performance of various priors in real data settings, using 5-fold cross-validation and computed the metrics MSPE, as defined in Section \ref{subsec:Prediction} for linear regression and ACC, specificity, sensitivity, AUC and F1 for logistic regression.  We also show how our method can be used to estimate the joint distribution of the regression coefficients to determine which groups of predictors are useful. 

{\bf Linear regression examples:}
The performance of the proposed method in linear regression setting was compared to HS and DL estimators utilizing two real datasets. First, The US communities and crime dataset, originally discussed by \cite{Redmond2002}, consists of 127 predictors that include socio-economic factors, alongside community and law enforcement-related variables. The dependent variable is the total number of violent crimes per 100,000 population. After
removing observations with missing values and non-numerical variables,  we retained 319 observations and 122 predictors for the analysis. This dataset can be found at \href{https://archive.ics.uci.edu/datasets}{UCI} machine learning repository. Second, The mouse gene expression dataset is originally from a study performed by \cite{Lan2006}, which aimed to identify regulatory networks and functional roles of genes associated with obesity and diabetes through the analysis of gene expression correlations across genetic dimensions. The dataset includes expression levels of 22,575 genes from 60 individuals (29 males and 31 females). Accompanying phenotypic data includes physiological measures such as number of fatty acid synthase, glycerol-3-phosphate acyltransferase (GPAT), and phosphoenopyruvate carboxykinase, measured in obese mouse populations using quantitative real-time PCR. We retrieved both gene expression and phenotypic data from the \href{(http://www.ncbi.nlm.nih.gov/geo}{Gene Expression Omnibus (GEO)} under accession number GSE3330, which has also been employed by \cite{Zhang2018} to evaluate performance of the DL prior in variable selection for sparse structures. For our analysis, we selected GPAT as the response variable and the top 1000 genes with the highest correlation to GPAT from the total set of 22,575 genes as candidate predictors.

The average MSPE for both datasets are shown in Table \ref{tab:Prediction_Crime&Mouse} along with the average difference in prediction performance (i.e., MSPE of other methods - MSPE of BECCA) and standard error of difference in parenthesis. The proposed method demonstrates the lowest prediction error in terms of MSPE compared to other evaluated approaches. However, The improvement in prediction performance provided by BECCA over HS and HS+ for the crime dataset is not statistically significant.

\begin{table}[htb]
\centering
\caption{Mean Squared Prediction Error (MSPE) computed using 5-fold cross validation for BECCA, HS, HS+ and DL priors on community crime and mouse genotype datasets. The improvement in prediction performance provided by BECCA over other techniques, measured as the MSPE of the other technique minus the MSPE of BECCA. In parenthese we give the average difference,  first entry and the standard error of the average difference, second entry in parentheses. Average differences of more than two standard deviations above zero are in bold. }
\vspace{3mm}
\label{tab:Prediction_Crime&Mouse}
\setlength\tabcolsep{12pt}
\begin{tabular}{@{}lcccc@{}} 
\toprule
 \textbf{Method} &  \textbf{BECCA} &  \textbf{HS} &  \textbf{HS+} &  \textbf{DL} \\ 
\midrule
Crime Data &  0.3795  & 0.3810 (0.003, 0.003) & 0.3819 (0.0.004, 0.004)  & 0.3985 ({\bf0.021}, 0.007)  \\
\midrule
Mouse Data &  0.6731 & 0.7082 ({\bf0.029}, 0.010) & 0.7173 ({\bf0.038}, 0.014) & 1.0391 ({\bf0.360}, 0.138)  \\
\bottomrule
\end{tabular}
\end{table}

\begin{figure}[htb]
\centering
\hspace*{-1.2cm} 
\includegraphics[width=1\linewidth]{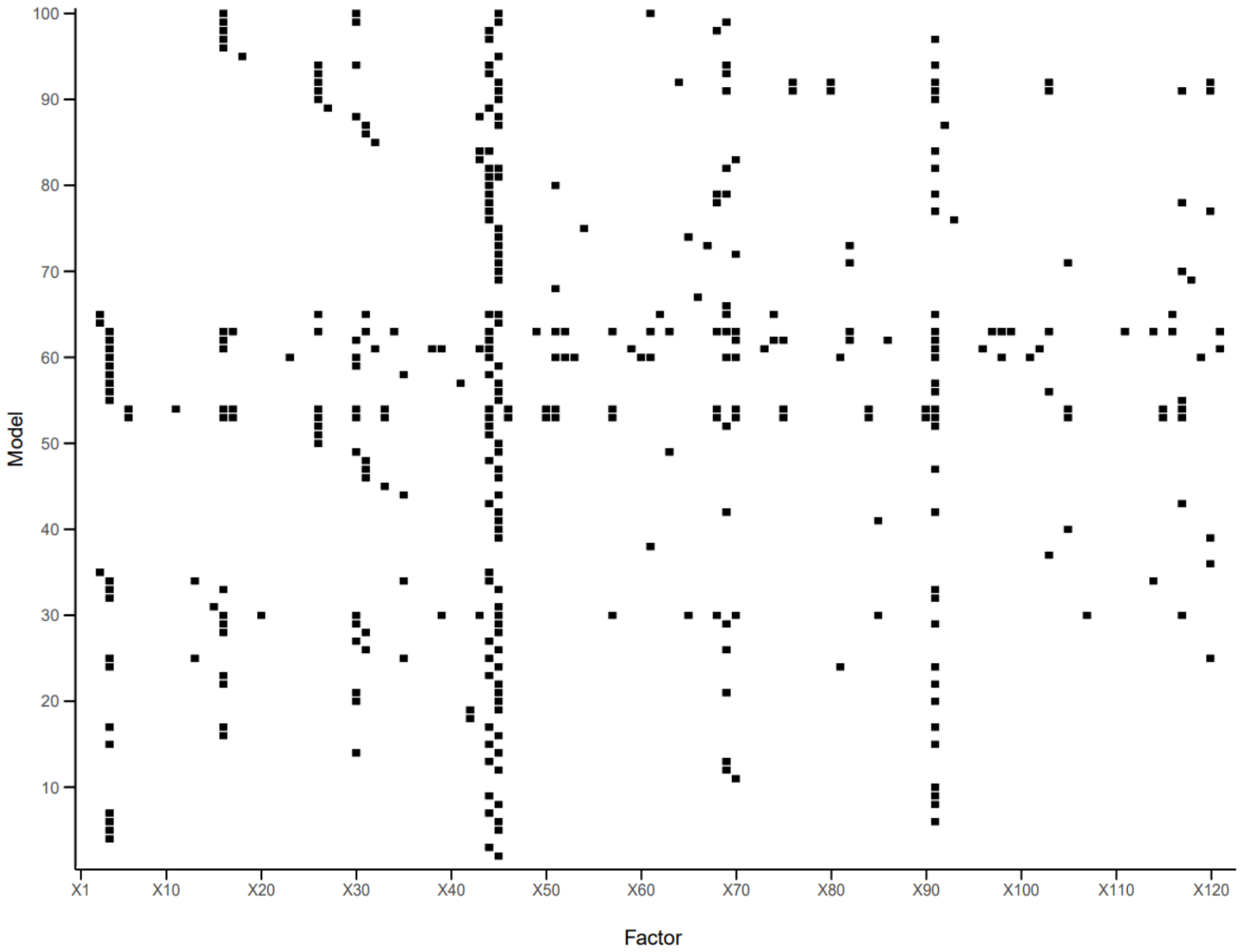}
\caption{\label{fig:Heat_linear} Factors in the top one hundred groups of predictors for the crime Dataset. Each row represent a model, or group of predictors, and each column represents factor. A black cell represents a given factor being included in a given model.}
\end{figure}

{\bf Logistic Regression examples:}
\label{subsec:LogitReg_real}
For logistic regression, we assessed the performance of different methods utilizing two binary classification datasets, the colon cancer dataset \citep{Alon1999} and breast cancer dataset \citep{William1993}. The \href{http://genomics-pubs.princeton.edu/oncology/}{colon cancer dataset}, originally consists of 2,000 gene expression levels derived from 62 tissue samples, out of which 40 are from tumor tissues and rest from normal colon tissues. Following \citep{Ma2024,Zhang2019}, we pre-processed the dataset (details in Supplementary Materials) and the final dataset consist with 387 predictor variables (gene expressions) for the analysis. The breast cancer dataset, publicly available at \href{https://archive.ics.uci.edu/dataset/17/breast+cancer+wisconsin+diagnostic}{UCI} repository, consists of 30 predictors describing characteristics of cell nuclei from 569 images, 212 of which are malignant, and the rest benign. Results of applying each method to colon and breast cancer datasets are provided in Table \ref{tab:Prediction_colone&breast}. In both datasets, the BECCA consistently performs better across most metrics. Even where it doesn't rank first, the difference is minimal, demonstrating competitive performance across all evaluation metrics.

{\bf Joint Inference for regression coefficients:}
In this section we show how our method can be used to determine which groups of predictors are useful predicting crime. Figure~\ref{fig:Heat_linear} presents a heatmap of the top 100 predictor combinations measured by the highest joint posterior probabilities, based on posterior samples of $\V\gamma$, which represents the inclusion probability for group of predictor variables. Rows represent predictor combinations, and columns indicate whether a variable is included (black) or excluded (white). Figure~\ref{fig:Heat_linear} shows three predictors ($X_{43}, X_{44}$ and $X_{90}$) consistently appear in the top combinations.  These predictors are (i) percentage of families headed by two parents, (ii) children in family housing with two parents, and (iii) the number of homeless people. This is consistent with previous findings \citep{Yerpude2017} and confirms the importance of these factors in U.S. community crime. The key to the factors used in this dataset is provided in the Supplementary for reference.
\begin{table}[htb]
\centering
\caption{Classification accuracy (ACC), Specificity (Spec), Sensitivity (Sens), AUC and F1 score for BECCA, HS, HS+ and DL priors on colon and breast cancer datasets. The highest metric values are bolded and the empirical standard errors are reported in parentheses.}
\vspace{3mm}
\label{tab:Prediction_colone&breast}
\setlength\tabcolsep{18pt} 
\begin{tabular}
{@{}cccccc@{}} 
\toprule
\textbf{Dataset} & \textbf{Metric} &  \textbf{BECCA} &  \textbf{HS} &  \textbf{HS+} &  \textbf{DL}   \\ 
\midrule
& ACC  &  \textbf{82.14} (9.29) &  78.81 (7.91) & 79.29 (10.09) &  75.48 (10.73)\\
\cmidrule(lr){2-6}
Colon & Spec  & \textbf{0.927} (0.101) & \textbf{0.927} (0.101) & 0.847 (0.166) &  \textbf{0.927} (0.101)\\
\cmidrule(lr){2-6}
cancer& Sens  & \textbf{0.783} (0.113) & 0.727 (0.117) & 0.769 (0.120) & 0.683 (0.155)\\
\cmidrule(lr){2-6}
& AUC  & 0.908 (0.087) & \textbf{0.909}(0.054) &  0.816 (0.111) & 0.882 (0.058)\\
\cmidrule(lr){2-6}
& F1 & \textbf{0.844} (0.083) & 0.810 (0.082) &  0.819 (0.104) & 0.775 (0.106)\\
\midrule
& ACC  &  \textbf{97.37} (1.31)  &  95.61 (2.07)& 94.55 (2.37) & 95.26 (2.55)\\
\cmidrule(lr){2-6}
Breast & Spec  & 0.945 (0.017) & \textbf{0.947} (0.029)& \textbf{0.947} (0.046) &  \textbf{0.947} (0.037)\\
\cmidrule(lr){2-6}
cancer& Sens  & \textbf{0.989} (0.019) & 0.960 (0.031) & 0.942 (0.030) & 0.954 (0.031)\\
\cmidrule(lr){2-6}
&AUC & \textbf{0.992} (0.003) & 0.991 (0.003) &  0.984 (0.008) & 0.986 (0.006)\\
\cmidrule(lr){2-6}
&F1 & \textbf{0.962} (0.019) & 0.941 (0.028) &  0.930 (0.024)& 0.937 (0.033)\\
\bottomrule 
\end{tabular}
\end{table} 
\clearpage
\section{DISCUSSION}
\label{sec:discussion}
In this paper, we introduced the BECCA prior,  a novel hierarchical prior for Bayesian variable selection, replacing the binary indicator variable in traditional spike-and-slab with a continuous Beta-distributed random variable that models the probability of predictor inclusion. This framework provides flexibility in modeling varying sparsity levels with the hyper-parameter introduced via a beta distribution while enabling coherent inference for both marginal and joint parameter spaces. In addition, it addresses the limitation of recent shrinkage priors (i.e., the lack of a coherent framework for variable and feature set inference), by allowing simultaneous shrinkage and variable selection. Moreover, the compatibility with Hamiltonian Monte Carlo (HMC) sampling ensures the computational efficiency through gradient-based methods. While the BECCA prior excels in high-dimensional settings, there is potential for further refinement, especially in its application to graphical and hierarchical models.

\section*{Acknowledgments}
The authors would like to acknowledge the support provided by the Australian Research Council (ARC) through the ARC Training Centre in Data Analytics for Resources and Environments (DARE) (Project IC190100031) and by the Human Technology Institute (HTI) at the University of Technology Sydney (UTS). 

\bibliographystyle{unsrtnat}
\bibliography{becca_arxiv}  
\appendix
\include{supplement}

\end{document}

%% file: supplement.tex
\section*{{\bf Supplementary Materials}: A {\bf Be}ta {\bf C}auchy-{\bf Ca}uchy (BECCA) shrinkage prior for Bayesian variable selection}
\section{Posterior computation}  
This section details the posterior computation processes for both the linear and logistic regression models, deriving the joint posterior distributions by combining the likelihood and the prior distributions.

The joint posterior for the linear regression model is given by,
{
\setlength\abovedisplayskip{10pt}
\setlength\belowdisplayskip{10pt}
\begin{align}
&\pi(\V \beta, \V \gamma, \sigma^2, g, u, v|\V y, \V X) \label{eq:posterior_linear}\\
&\propto f(\V y| \V \beta, \V \gamma, \sigma^2, \V X) \pi(\V \beta|\V \gamma, \sigma^2, g) \pi(\V \gamma| u,v) \pi(\sigma^2) \pi(g,u,v)  \notag\\
&\propto \frac{(1+g^2)^{-1}(\sigma^2)^{-a-1} e^{-\frac{b}{\sigma^2}}}{(1+u^2)(1+v^2)} \times \frac{1}{\sigma^2}e^{- \frac{1}{\sigma^2} (\V y - \V X \V \beta)^\top (\V y - \V X \V \beta)}  \notag\\
&\times \left[\frac{\Gamma(2u)}{\sqrt{g\sigma^2}\Gamma(u^2)} \right]^p \prod_{j=1}^{p} \frac{1}{\gamma_j} e^{ -\frac{\beta_j^2}{2g\sigma^2\gamma_j^2} } \times \V{\gamma}_j^{u-1} (1-\V{\gamma}_j)^{v-1}. \notag
\end{align}
}

Similarly, the joint posterior for the logistic regression model is given by,
{
\setlength\abovedisplayskip{6pt}
\setlength\belowdisplayskip{6pt}
\begin{align}
&\pi(\V \beta, \V \gamma, g, u, v|\V y, \V X) 
\label{eq:posterior_logit}\\
&\propto f(\V y| \V \beta, \V \gamma, \V X) \pi(\V \beta|\V \gamma, g) \pi(\V \gamma| u,v) \pi(g,u,v)  \notag\\
&\propto \frac{(1+g^2)^{-1}}{(1+u^2)(1+v^2)} \prod_{i=1}^{n} \left[\frac{1}{1 + e^{-\V X_i^\top \V \beta}}\right]^{y_i}\left[\frac{1}{1 + e^{\V X_i^\top \V \beta}}\right]^{1 - y_i} \notag\\
&\quad \times \left[\frac{\Gamma(2u)}{\Gamma(u^2)\sqrt{g}} \right]^p \prod_{j=1}^{p} \frac{1}{\gamma_j} e^{ -\frac{\beta_j^2}{2g\gamma_j^2} } \V{\gamma}_j^{u-1} (1-\V{\gamma}_j)^{v-1}.  \notag
\end{align}
}
\clearpage
\section{Simulation study: Data generation and model training}
In this section, we provide a comprehensive description of the data generation process and model training procedures for both linear and logistic regression simulations. 

For linear regression, the predictor vectors $\mathbf{X}_i$ for $i = 1,...,n$ were independently drawn from $\mathcal{N}_k(\mathbf{0},\Sigma)$ with fixed correlations among predictors, where $\text{diag}(\Sigma) = \mathbf{1}$ and $\Sigma_{ij} = 0.75$ for $i \neq j$ and all active regression coefficients $(\beta_j)$ were generated from $\mathcal{N}(0,c.\gamma_j)$ and remaining were set to zero following the Algorithm \ref{alg:data_generation_linear}. For logistic regression, we adopted the auto-regressive correlation design, where $\Sigma_{ij} = 0.65^{|i-j|}$ for $i \neq j$ capturing varying correlations among predictors and all active regression coefficient $(\beta_j)$ were generated from $\text{Unif}(2,7.5)$ following the Algorithm \ref{alg:data_generation_logistic}.

\begin{figure}[htb]
\centering

\begin{subfigure}[t]{0.48\linewidth}
    \centering
    \includegraphics[width=\linewidth]{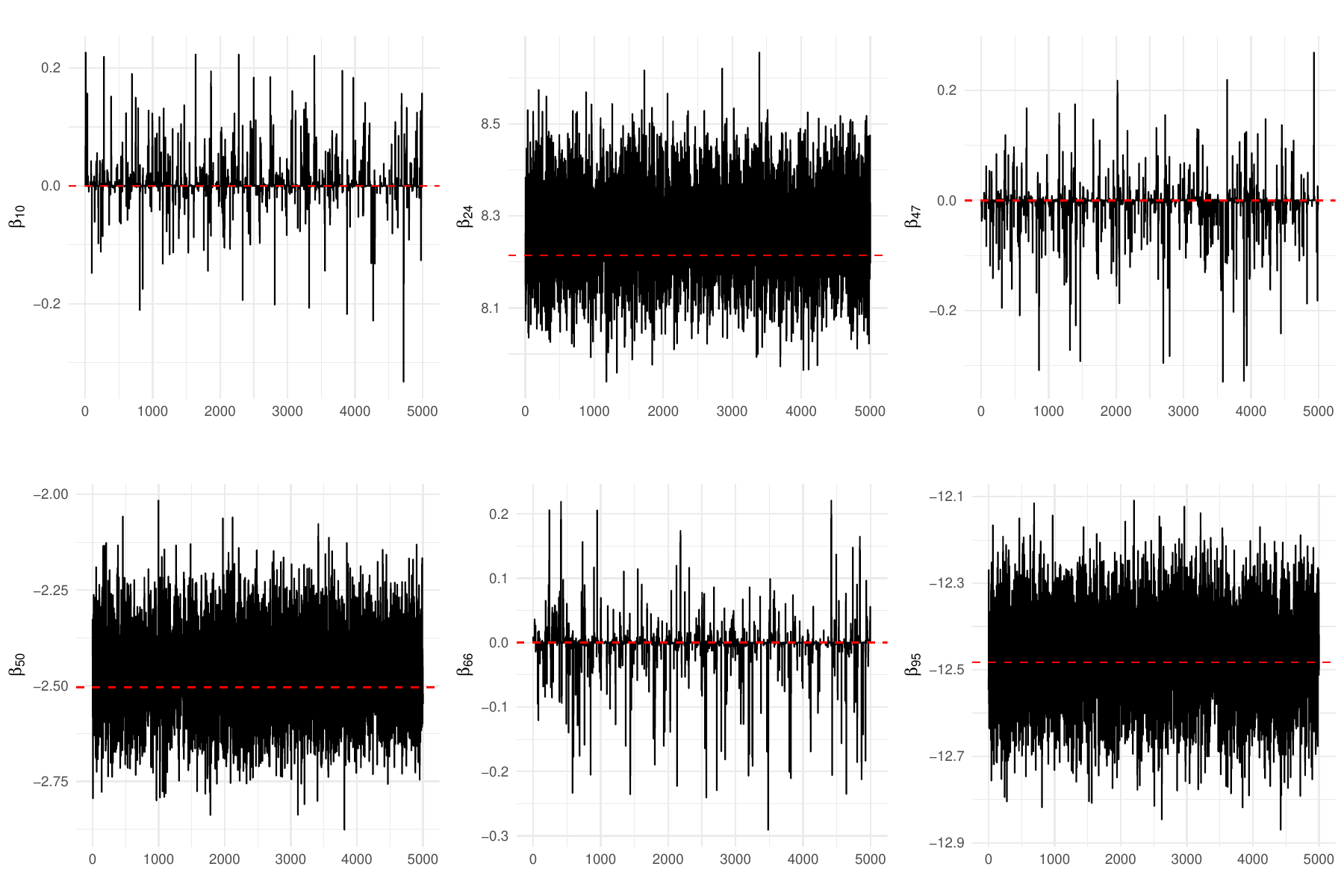}
    \caption{}
    \label{fig:Trace_linearX100}
\end{subfigure}
\hfill
\begin{subfigure}[t]{0.48\linewidth}
    \centering
    \includegraphics[width=\linewidth]{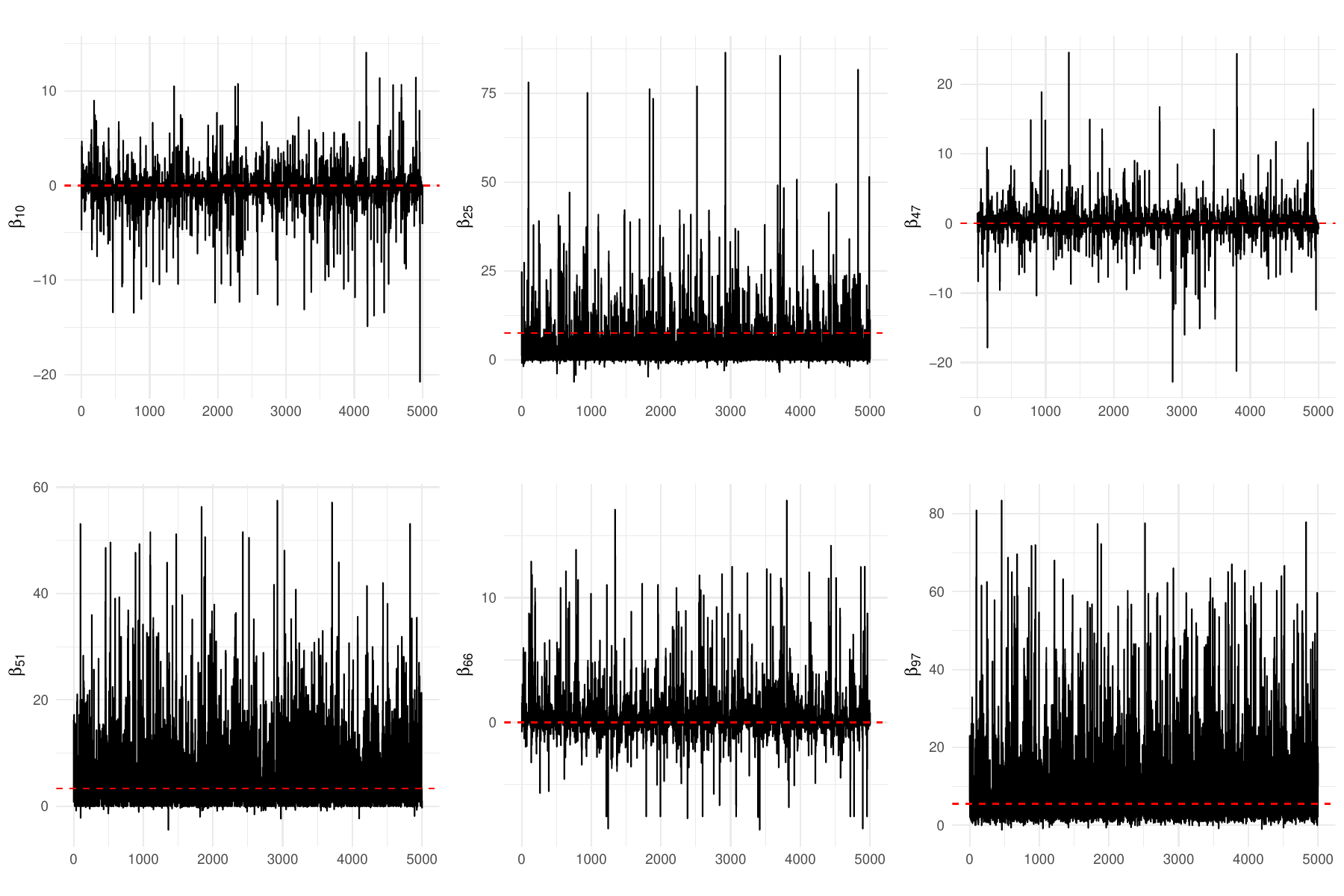}
    \caption{}
    \label{fig:Trace_logitX100}
\end{subfigure}

\vspace{.2in}
\caption{Trace plots generated under BECCA prior applied to (a) linear and (b) logistic regression with a simulation setting of $n=100$ and $p=100$. The red dashed line shows the true coefficient value.}
\end{figure}

\begin{figure*}[htb]
\centering
\begin{subfigure}[t]{0.45\textwidth}
        \centering
\includegraphics[width=1\linewidth]{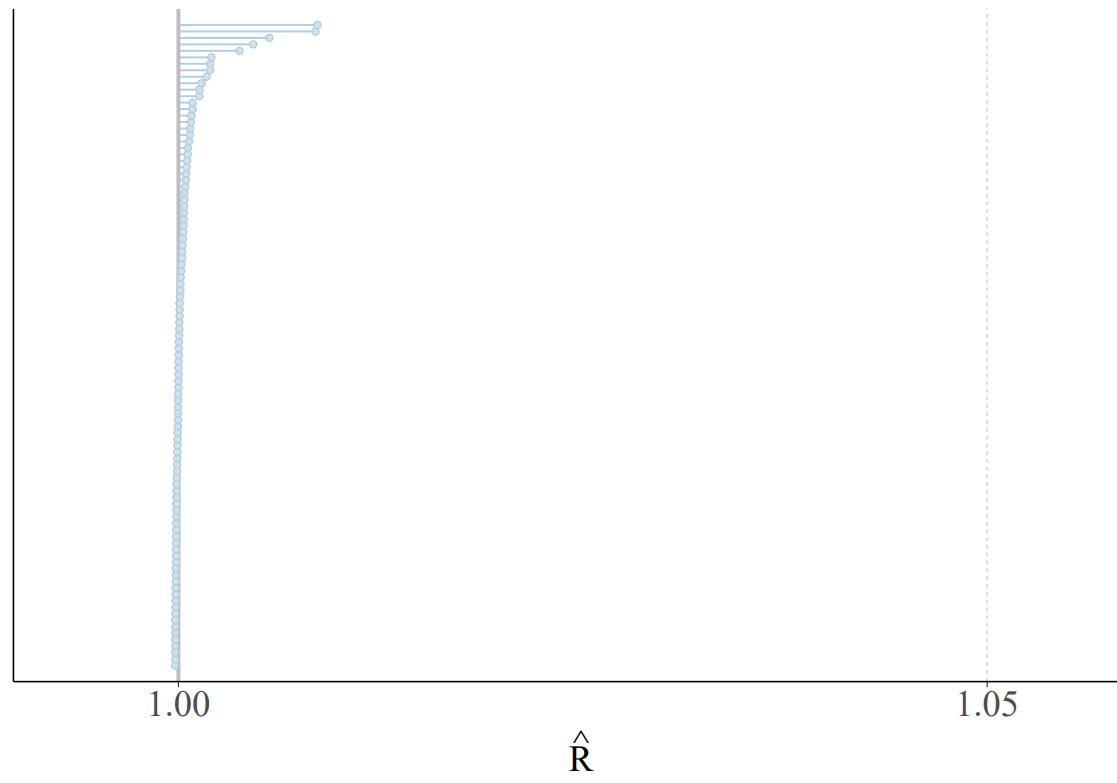}
\caption{}
\end{subfigure}
\begin{subfigure}[t]{0.45\textwidth}
        \centering
\includegraphics[width=1\linewidth]{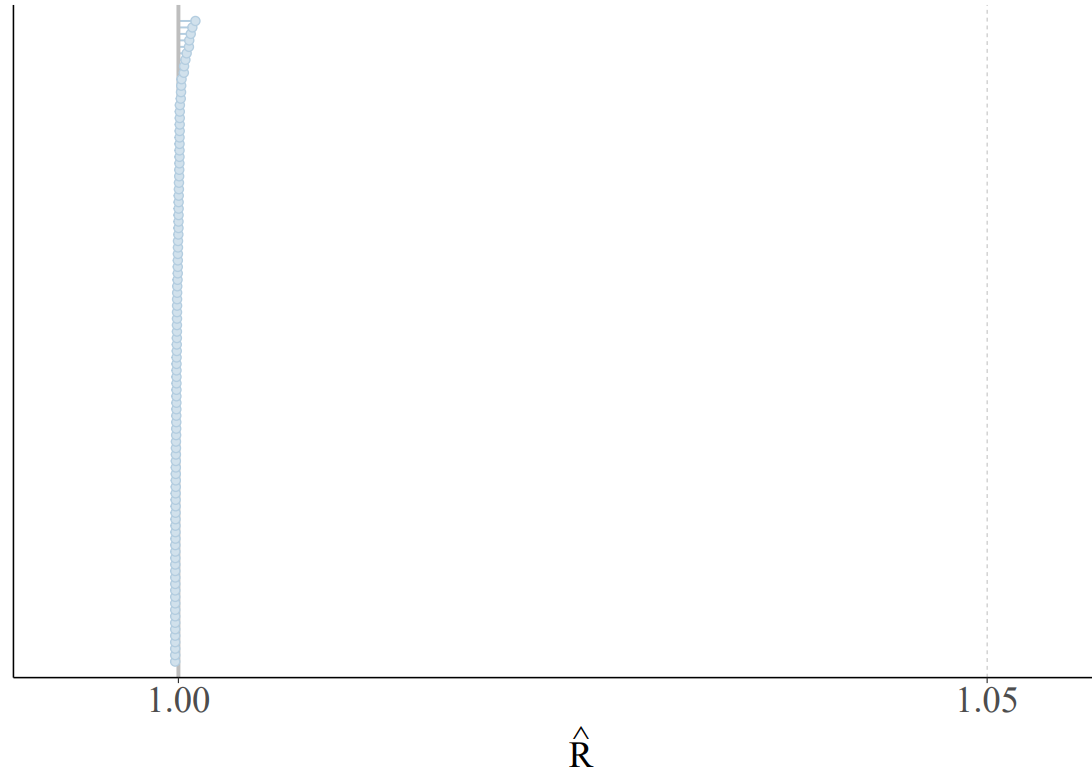}
\caption{}
\end{subfigure}
\begin{subfigure}[t]{0.45\textwidth}
        \centering
\includegraphics[width=1\linewidth]{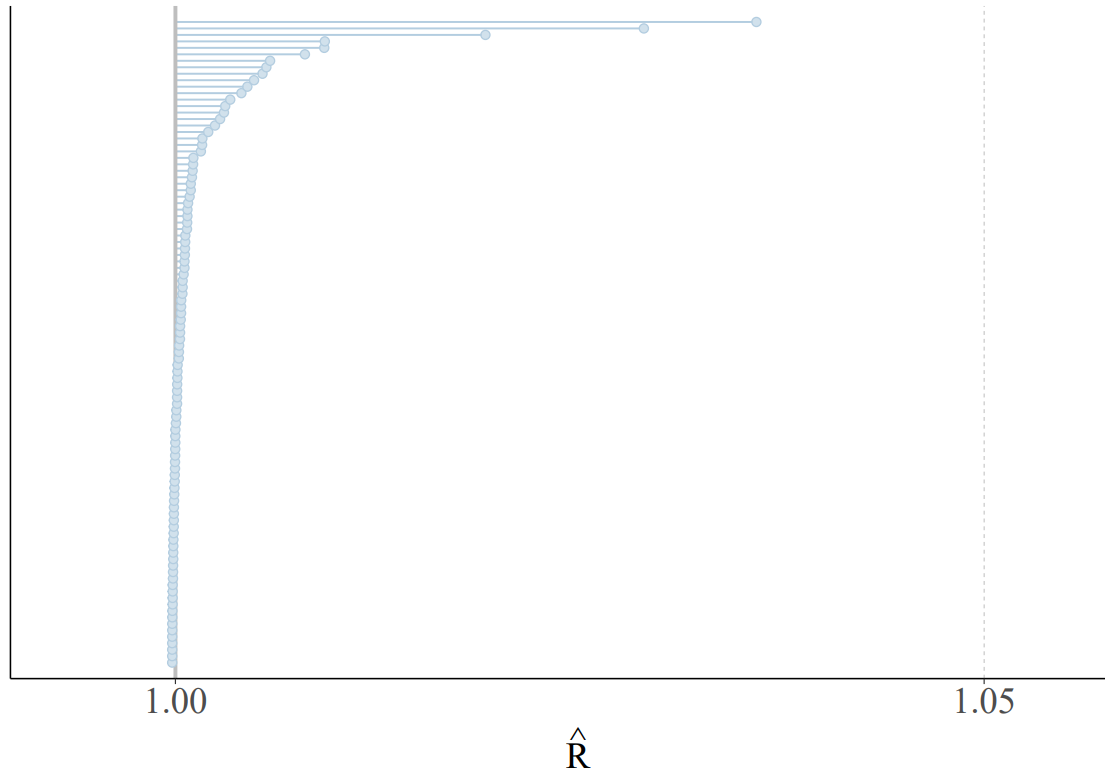}
\caption{}
\end{subfigure}
\begin{subfigure}[t]{0.45\textwidth}
        \centering
\includegraphics[width=1\linewidth]{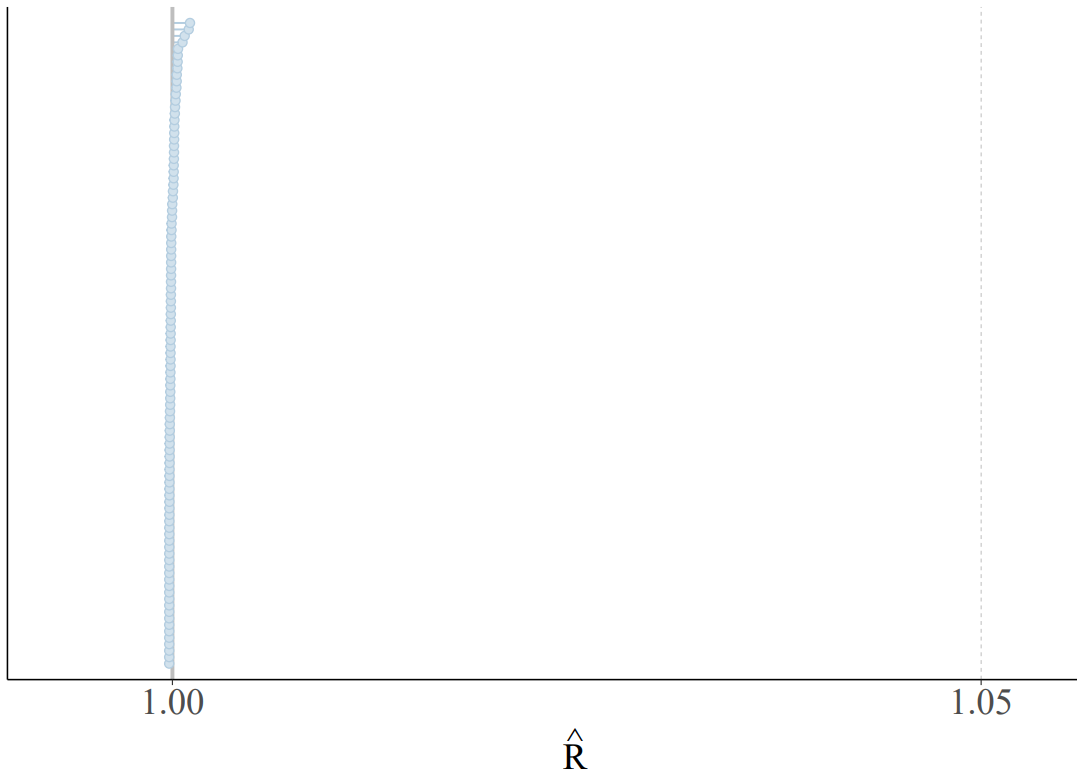}
\caption{}
\end{subfigure}
\caption{Gelman-Rubin statistics ($\hat{R}$) for each linear regression coefficient in the MCMC chains generated under (a)  BECCA, (b) HS , (c) HS+ and (d) DL priors.}
\label{fig:Rhat_linear}
\end{figure*}

\begin{figure*}[htb]
\centering
\begin{subfigure}[t]{0.45\textwidth}
        \centering
\includegraphics[width=1\linewidth]{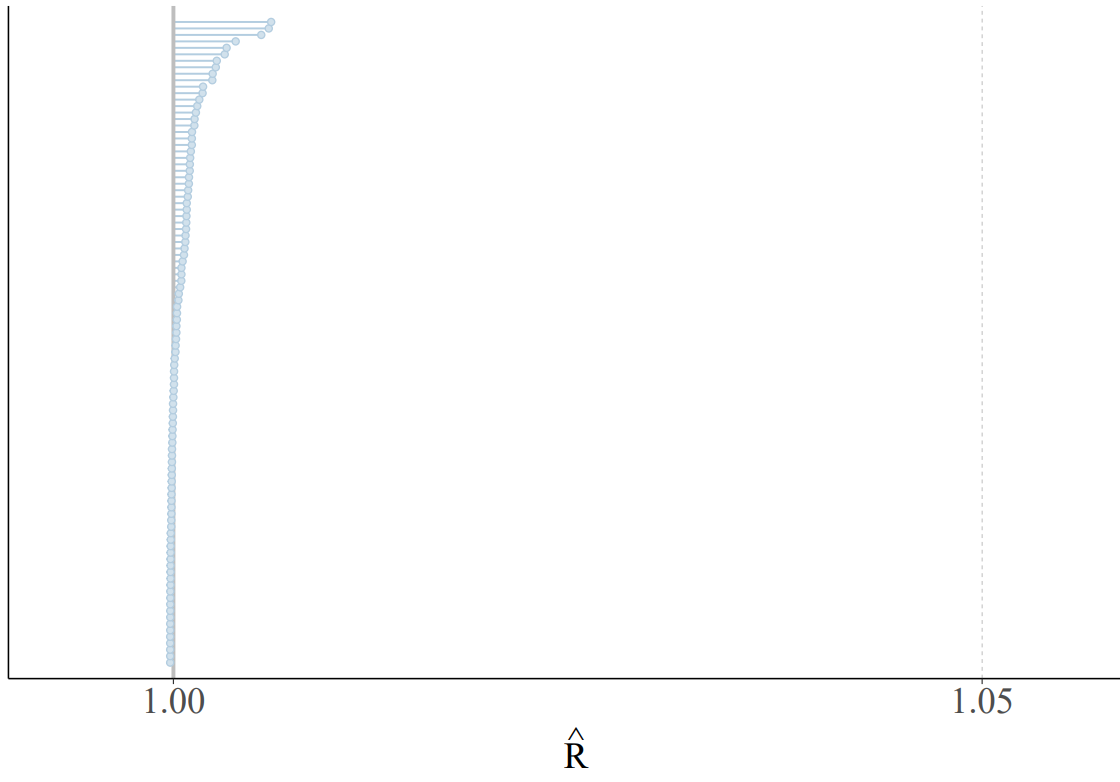}
\caption{}
\end{subfigure}
\begin{subfigure}[t]{0.45\textwidth}
        \centering
\includegraphics[width=1\linewidth]{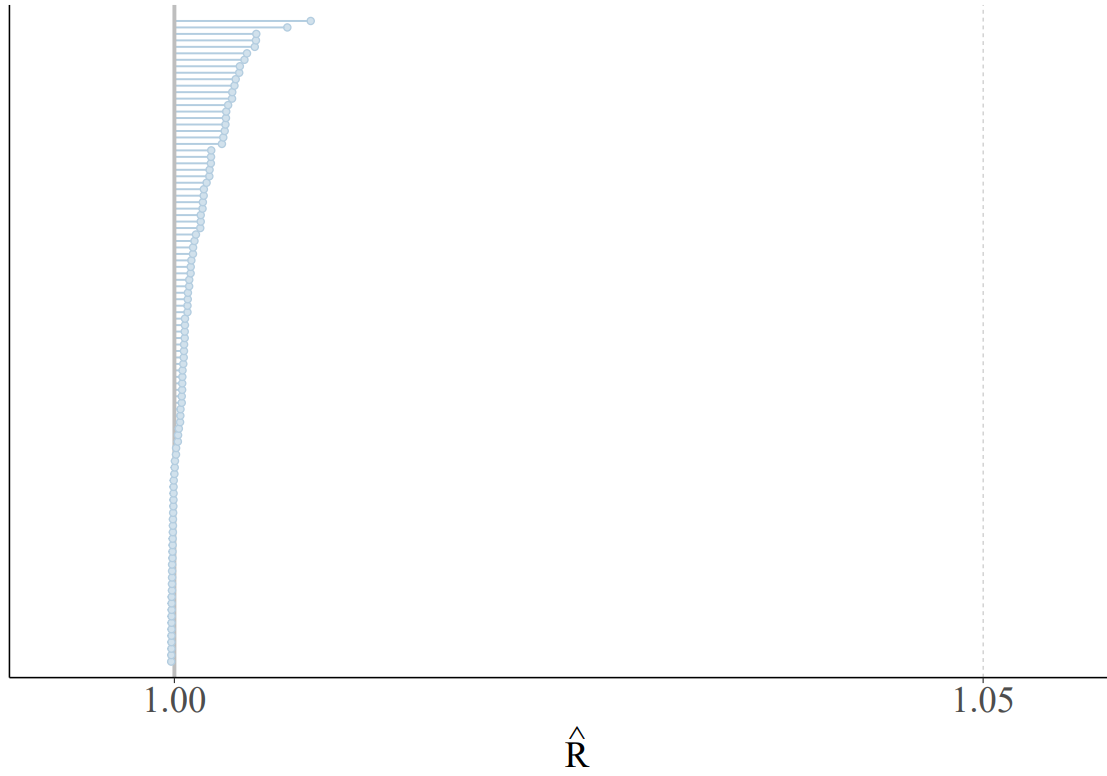}
\caption{}
\end{subfigure}
\begin{subfigure}[t]{0.45\textwidth}
        \centering
\includegraphics[width=1\linewidth]{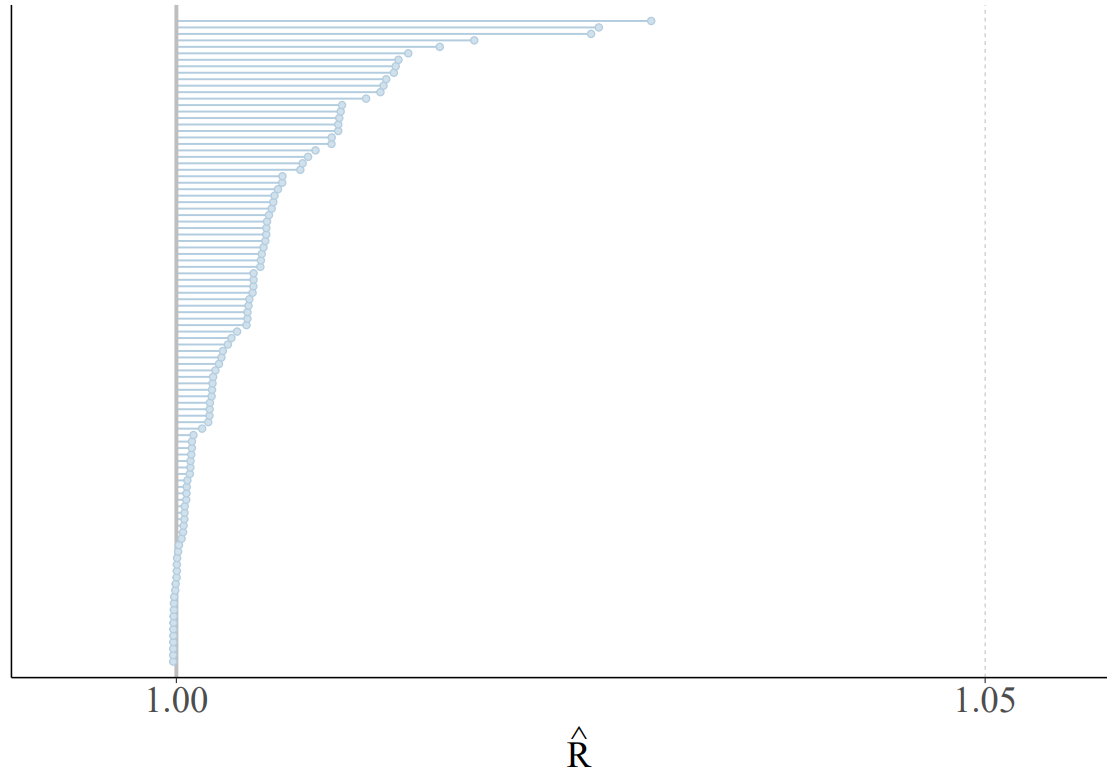}
\caption{}
\end{subfigure}
\begin{subfigure}[t]{0.45\textwidth}
        \centering
\includegraphics[width=1\linewidth]{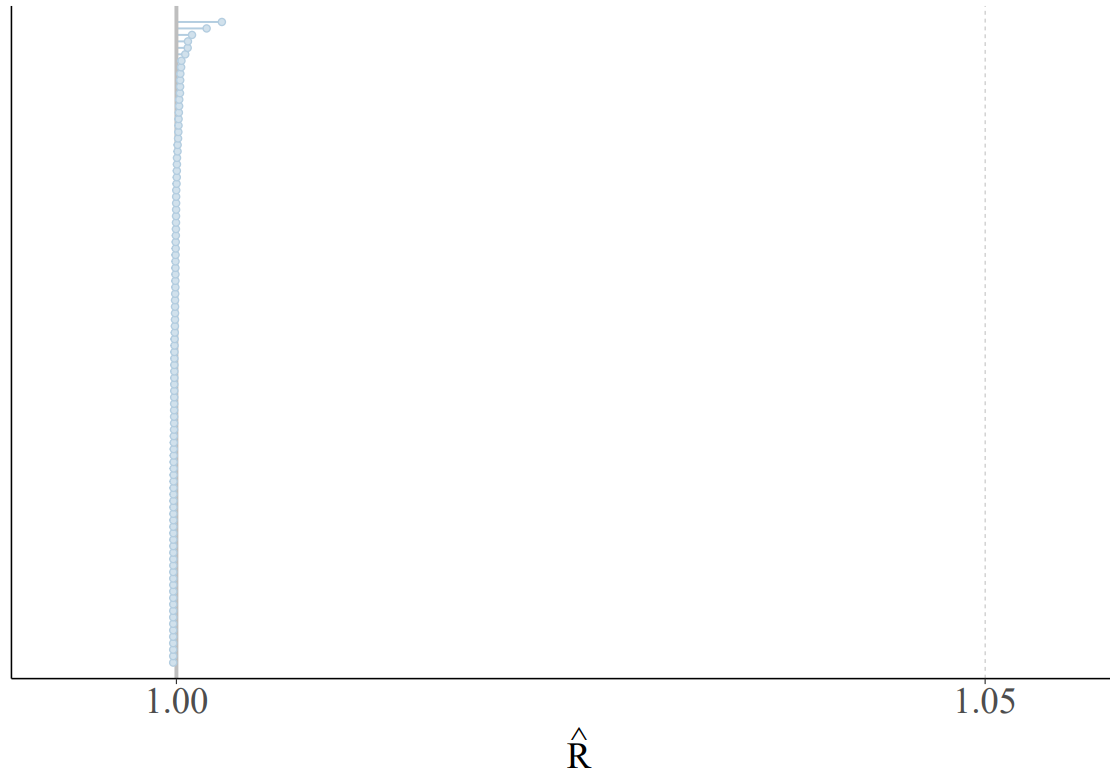}
\caption{}
\end{subfigure}
\caption{Gelman-Rubin statistics ($\hat{R}$) for each logistic regression coefficient in the MCMC chains generated under (a)  BECCA, (b) HS , (c) HS+ and (d) DL priors.}
\label{fig:Rhat_logit}
\end{figure*}

For each simulation setting, a total of 10,000 samples were drawn, with the initial 5,000 samples discarded as burn-in to ensure stabilization. Each simulation scenario was replicated 50 times, and the results were summarized by averaging each performance metric across these replicates. We assessed MCMC convergence both visually and numerically. Figures \ref{fig:Trace_linearX100}
 and \ref{fig:Trace_logitX100} show the trace plots of selected linear and logistic regression coefficients, respectively. Additionally, the Gelman-Rubin statistic ($\hat{R}$) was calculated for the regression coefficients to numerically assess convergence under different methods, and the results are displayed in Figure \ref{fig:Rhat_linear} for linear regression and Figure \ref{fig:Rhat_logit} for logistic regression respectively.

\begin{algorithm}[htb]
\caption{Generate Data for Bayesian Linear Regression}
\label{alg:data_generation_linear} 
\begin{algorithmic}[1]  
\State \textbf{Input:} Number of observations $N$, total number of predictors $p$, number of active predictors $p_1$
\State \textbf{Output:} Data matrix $\mathbf X$, response vector $\mathbf y$, regression coefficients $\boldsymbol \beta$
\vspace{3mm}
\State \textbf{Generate predictor variables $\mathbf{X}$:} 
Generate $\mathbf{X}_{N\times p}$ from $\textit{MVN}(\mathbf{0}, \Sigma)$; $\Sigma_{ij} = 0.75$ $\forall i \neq j$ 

\vspace{1mm}
    \Statex \textbf{Generate regression coefficients $\boldsymbol{\beta}$:}
    \Statex \hspace{5mm} Set $\sigma^2 =1, g=N$ 
    \Statex \hspace{5mm} Set $\boldsymbol \gamma_{\mathbf A_1} = \text{rep}(1, p_1)$ and $\V \gamma_{\mathbf A_0} = \text{rep}(0, p - p_1))$
    \Statex \hspace{5mm}Generate $\beta_j$ from $\textit{N}(0, g\sigma^2\gamma_j^2);$ $\forall j \in \mathbf A_1$ and $\beta_j = 0;$ $\forall j \in \mathbf A_0$
\vspace{1mm}
\State \textbf{Generate response variable $\mathbf{y}$:}
\For{$i = 1$ to $N$}
    \State Generate $y_{[i]}$ from $\text{N}(\mathbf{x}_{[i]}\boldsymbol{\beta}, \sigma^2)$
\EndFor
\vspace{3mm}
\State \textbf{return} $\mathbf{X}, \mathbf{y}, \boldsymbol{\beta}$
\end{algorithmic}
\end{algorithm}

\begin{algorithm}[htb]
\caption{Generate Data for Bayesian Logistic Regression}
\label{alg:data_generation_logistic} 
\begin{algorithmic}[1]  
\State \textbf{Input:} Number of observations $N$, total number of predictors $p$, number of active predictors $p_1$
\State \textbf{Output:} Data matrix $\mathbf X$, response vector $\mathbf y$, regression coefficients $\boldsymbol \beta$

\State \textbf{Generate predictor variables $\mathbf{X}$:} 
Generate $\mathbf{X}_{N\times p}$ from $\textit{MVN}(\mathbf{0}, \Sigma)$; $\Sigma_{ij} = 0.65^{|i-j|}$ $\forall i \neq j$ 

\vspace{1mm}
\State \textbf{Generate regression coefficients $\boldsymbol{\beta}$:}
    \State \hspace{5mm} Set $\V \gamma_{\V A_1} = \text{rep}(1, p_1)$ and $\V \gamma_{\V A_0} = \text{rep}(0, p - p_1))$
    \State \hspace{5mm}Generate $\beta_j$ from $\text{Unif}(2, 7.5);$ $\forall j \in \V A_1$ and $\beta_j = 0;$ $\forall j \in \V A_0$
\vspace{1mm}
\State \textbf{Generate response variable $\mathbf{y}$:}
\For{$i = 1$ to $N$}
    \State Generate $y_{[i]}$ from $\text{Bernoulli}(\text{logit}^{-1}(\mathbf{x}_{[i]}\boldsymbol{\beta}))$
\EndFor
\vspace{3mm}
\State \textbf{return} $\mathbf{X}, \mathbf{y}, \boldsymbol{\beta}$
\end{algorithmic}
\end{algorithm}
\clearpage
\section{Real data applications}
\subsection{Data pre-processing and model training}
\label{appendix_real_data}
In this section, we outline the data pre-processing steps and model training procedures for real-world datasets, particularly, the colon cancer and US crime datasets.

Following the approach of our simulation study, we trained the models using HMC with the NUTS algorithm, performing 10,000 iterations, with the first 5,000 discarded as burn-in. The hyperparameters $a$ and $b$ for the inverse gamma prior of the variance parameter $\sigma^2$ were both set to 0.5 based on common practices. Before implementing different methods, all datasets were standardized to have zero mean and unit variance.

For the colon dataset, we applied a log10 transformation to the gene expression data and filtered out variables with no significant difference between tumor and normal tissues using the Wilcoxon rank-sum test at a 0.05 significance level. The final dataset contained 387 predictor variables (gene expressions) across 62 tissue samples. The US crime dataset initially had 127 predictors, including socioeconomic and community-related variables. After removing observations with missing values and non-numerical variables, we retained 319 observations and 122 predictors for the analysis. No specific data pre-processing was applied to the mouse or breast cancer datasets, which were used in their original form.

\subsection{Key to factors in crime dataset}
\begin{longtable}{p{0.075\textwidth}p{0.24\textwidth}p{0.6\textwidth}} 
\toprule
\textbf{Factor} & \textbf{Factor Name} & \textbf{Description} \\ 
\midrule
\endfirsthead
\toprule
\textbf{Factor} & \textbf{Factor Name} & \textbf{Description} \\ 
\midrule
\endhead
\midrule
\multicolumn{3}{r}{\textit{Continued on next page}} \\
\midrule
\endfoot
\bottomrule
\endlastfoot
X1 & householdsize & Mean people per household (numeric - decimal) \\
X2 & racepctblack & Percentage of population that is African American (numeric - decimal) \\
X3 & racePctWhite & Percentage of population that is Caucasian (numeric - decimal) \\
X4 & racePctAsian & Percentage of population that is of Asian heritage (numeric - decimal) \\
X5 & racePctHisp & Percentage of population that is of Hispanic heritage (numeric - decimal) \\
X6 & agePct12t21 & Percentage of population aged 12–21 (numeric - decimal) \\
X7 & agePct12t29 & Percentage of population aged 12–29 (numeric - decimal) \\
X8 & agePct16t24 & Percentage of population aged 16–24 (numeric - decimal) \\
X9 & agePct65up & Percentage of population aged 65 and over (numeric - decimal) \\
X10 & numbUrban & Number of people living in urban areas (numeric - decimal) \\
X11 & pctUrban & Percentage of people living in urban areas (numeric - decimal) \\
X12 & medIncome & Median household income (numeric - decimal) \\
X13 & pctWWage & Percentage of households with wage or salary income in 1989 (numeric - decimal) \\
X14 & pctWFarmSelf & Percentage of households with farm or self-employment income in 1989 (numeric - decimal) \\
X15 & pctWInvInc & Percentage of households with investment/rent income in 1989 (numeric - decimal) \\
X16 & pctWSocSec & Percentage of households with social security income in 1989 (numeric - decimal) \\
X17 & pctWPubAsst & Percentage of households with public assistance income in 1989 (numeric - decimal) \\
X18 & pctWRetire & Percentage of households with retirement income in 1989 (numeric - decimal) \\
X19 & medFamInc & Median family income (numeric - decimal) \\
X20 & perCapInc & Per capita income (numeric - decimal) \\
X21 & whitePerCap & Per capita income for Caucasians (numeric - decimal) \\
X22 & blackPerCap & Per capita income for African Americans (numeric - decimal) \\
X23 & indianPerCap & Per capita income for Native Americans (numeric - decimal) \\
X24 & AsianPerCap & Per capita income for people with Asian heritage (numeric - decimal) \\
X25 & OtherPerCap & Per capita income for people with "other" heritage (numeric - decimal) \\
X26 & HispPerCap & Per capita income for people with Hispanic heritage (numeric - decimal) \\
X27 & NumUnderPov & Number of people under the poverty level (numeric - decimal) \\
X28 & PctPopUnderPov & Percentage of people under the poverty level (numeric - decimal) \\
X29 & PctLess9thGrade & Percentage of people 25 and over with less than a 9th-grade education (numeric - decimal) \\
X30 & PctNotHSGrad & Percentage of people 25 and over who are not high school graduates (numeric - decimal) \\
X31 & PctBSorMore & Percentage of people 25 and over with a bachelor's degree or higher education (numeric - decimal) \\
X32 & PctUnemployed & Percentage of people 16 and over, in the labor force, and unemployed (numeric - decimal) \\
X33 & PctEmploy & Percentage of people 16 and over who are employed (numeric - decimal) \\
X34 & PctEmplManu & Percentage of people 16 and over who are employed in manufacturing (numeric - decimal) \\
X35 & PctEmplProfServ & Percentage of people 16 and over who are employed in professional services (numeric - decimal) \\
X36 & PctOccupManu & Percentage of people 16 and over who are employed in manufacturing (numeric - decimal) \\
X37 & PctOccupMgmtProf & Percentage of people 16 and over who are employed in management or professional occupations (numeric - decimal) \\
X38 & MalePctDivorce & Percentage of males who are divorced (numeric - decimal) \\
X39 & MalePctNevMarr & Percentage of males who have never married (numeric - decimal) \\
X40 & FemalePctDiv & Percentage of females who are divorced (numeric - decimal) \\
X41 & TotalPctDiv & Percentage of the population who are divorced (numeric - decimal) \\
X42 & PersPerFam & Mean number of people per family (numeric - decimal) \\
X43 & PctFam2Par & Percentage of families (with kids) that are headed by two parents (numeric - decimal) \\
X44 & PctKids2Par & Percentage of kids in family housing with two parents (numeric - decimal) \\
X45 & PctYoungKids2Par & Percentage of kids aged 4 and under in two-parent households (numeric - decimal) \\
X46 & PctTeen2Par & Percentage of kids aged 12–17 in two-parent households (numeric - decimal) \\
X47 & PctWorkMomYoungKids & Percentage of moms of kids aged 6 and under in the labor force (numeric - decimal) \\
X48 & PctWorkMom & Percentage of moms of kids under 18 in the labor force (numeric - decimal) \\
X49 & NumIlleg & Number of kids born to never-married mothers (numeric - decimal) \\
X50 & PctIlleg & Percentage of kids born to never-married mothers (numeric - decimal) \\
X51 & NumImmig & Total number of people known to be foreign born (numeric - decimal) \\
X52 & PctImmigRecent & Percentage of immigrants who immigrated within the last 3 years (numeric - decimal) \\
X53 & PctImmigRec5 & Percentage of immigrants who immigrated within the last 5 years (numeric - decimal) \\
X54 & PctImmigRec8 & Percentage of immigrants who immigrated within the last 8 years (numeric - decimal) \\
X55 & PctImmigRec10 & Percentage of immigrants who immigrated within the last 10 years (numeric - decimal) \\
X56 & PctRecentImmig & Percentage of the population who have immigrated within the last 3 years (numeric - decimal) \\
X57 & PctRecImmig5 & Percentage of the population who have immigrated within the last 5 years (numeric - decimal) \\
X58 & PctRecImmig8 & Percentage of the population who have immigrated within the last 8 years (numeric - decimal) \\
X59 & PctRecImmig10 & Percentage of the population who have immigrated within the last 10 years (numeric - decimal) \\
X60 & PctSpeakEnglOnly & Percentage of people who speak only English (numeric - decimal) \\
X61 & PctNotSpeakEnglWell & Percentage of people who do not speak English well (numeric - decimal) \\
X62 & PctLargHouseFam & Percentage of family households that are large (6 or more) (numeric - decimal) \\
X63 & PctLargHouseOccup & Percentage of all occupied households that are large (6 or more people) (numeric - decimal) \\
X64 & PersPerOccupHous & Mean persons per household (numeric - decimal) \\
X65 & PersPerOwnOccHous & Mean persons per owner-occupied household (numeric - decimal) \\
X66 & PersPerRentOccHous & Mean persons per rental household (numeric - decimal) \\
X67 & PctPersOwnOccup & Percentage of people in owner-occupied households (numeric - decimal) \\
X68 & PctPersDenseHous & Percentage of persons in dense housing (more than 1 person per room) (numeric - decimal) \\
X69 & PctHousLess3BR & Percentage of housing units with less than 3 bedrooms (numeric - decimal) \\
X70 & MedNumBR & Median number of bedrooms (numeric - decimal) \\
X71 & HousVacant & Number of vacant households (numeric - decimal) \\
X72 & PctHousOccup & Percent of housing occupied (numeric - decimal) \\
X73 & PctHousOwnOcc & Percent of households owner occupied (numeric - decimal) \\
X74 & PctVacantBoarded & Percent of vacant housing that is boarded up (numeric - decimal) \\
X75 & PctVacMore6Mos & Percent of vacant housing that has been vacant more than 6 months (numeric - decimal) \\
X76 & MedYrHousBuilt & Median year housing units built (numeric - decimal) \\
X77 & PctHousNoPhone & Percent of occupied housing units without phone (numeric - decimal) \\
X78 & PctWOFullPlumb & Percent of housing without complete plumbing facilities (numeric - decimal) \\
X79 & OwnOccLowQuart & Owner occupied housing - lower quartile value (numeric - decimal) \\
X80 & OwnOccMedVal & Owner occupied housing - median value (numeric - decimal) \\
X81 & OwnOccHiQuart & Owner occupied housing - upper quartile value (numeric - decimal) \\
X82 & RentLowQ & Rental housing - lower quartile rent (numeric - decimal) \\
X83 & RentMedian & Rental housing - median rent (numeric - decimal) \\
X84 & RentHighQ & Rental housing - upper quartile rent (numeric - decimal) \\
X85 & MedRent & Median gross rent (numeric - decimal) \\
X86 & MedRentPctHousInc & Median gross rent as a percentage of household income (numeric - decimal) \\
X87 & MedOwnCostPctInc & Median owner's cost as a percentage of household income (numeric - decimal) \\
X88 & MedOwnCostPctIncNoMtg & Median owner's cost as a percentage of household income - for owners without a mortgage (numeric - decimal) \\
X89 & NumInShelters & Number of people in homeless shelters (numeric - decimal) \\
X90 & NumStreet & Number of homeless people counted in the street (numeric - decimal) \\
X91 & PctForeignBorn & Percent of people foreign born (numeric - decimal) \\
X92 & PctBornSameState & Percent of people born in the same state as currently living (numeric - decimal) \\
X93 & PctSameHouse85 & Percent of people living in the same house as in 1985 (numeric - decimal) \\
X94 & PctSameCity85 & Percent of people living in the same city as in 1985 (numeric - decimal) \\
X95 & PctSameState85 & Percent of people living in the same state as in 1985 (numeric - decimal) \\
X96 & LemasSwornFT & Number of sworn full-time police officers (numeric - decimal) \\
X97 & LemasSwFTPerPop & Sworn full-time police officers per 100K population (numeric - decimal) \\
X98 & LemasSwFTFieldOps & Number of sworn full-time police officers in field operations (numeric - decimal) \\
X99 & LemasSwFTFieldPerPop & Sworn full-time police officers in field operations per 100K population (numeric - decimal) \\
X100 & LemasTotalReq & Total requests for police (numeric - decimal) \\
X101 & LemasTotReqPerPop & Total requests for police per 100K population (numeric - decimal) \\
X102 & PolicReqPerOffic & Total requests for police per police officer (numeric - decimal) \\
X103 & PolicPerPop & Police officers per 100K population (numeric - decimal) \\
X104 & RacialMatchCommPol & Racial match between the community and the police force (numeric - decimal) \\
X105 & PctPolicWhite & Percent of police that are Caucasian (numeric - decimal) \\
X106 & PctPolicBlack & Percent of police that are African American (numeric - decimal) \\
X107 & PctPolicHisp & Percent of police that are Hispanic (numeric - decimal) \\
X108 & PctPolicAsian & Percent of police that are Asian (numeric - decimal) \\
X109 & PctPolicMinor & Percent of police that are minorities (numeric - decimal) \\
X110 & OfficAssgnDrugUnits & Number of officers assigned to special drug units (numeric - decimal) \\
X111 & NumKindsDrugsSeiz & Number of different kinds of drugs seized (numeric - decimal) \\
X112 & PolicAveOTWorked & Police average overtime worked (numeric - decimal) \\
X113 & LandArea & Land area in square miles (numeric - decimal) \\
X114 & PopDens & Population density in persons per square mile (numeric - decimal) \\
X115 & PctUsePubTrans & Percent of people using public transit for commuting (numeric - decimal) \\
X116 & PolicCars & Number of police cars (numeric - decimal) \\
X117 & PolicOperBudg & Police operating budget (numeric - decimal) \\
X118 & LemasPctPolicOnPatr & Percent of sworn full-time police officers on patrol (numeric - decimal) \\
X119 & LemasGangUnitDeploy & Gang unit deployed (numeric - decimal - ordinal: 0 = NO, 1 = YES, 0.5 = Part Time) \\
X120 & LemasPctOfficDrugUn & Percent of officers assigned to drug units (numeric - decimal) \\
X121 & PolicBudgPerPop & Police operating budget per population (numeric - decimal) \\
X122 & ViolentCrimesPerPop & Total number of violent crimes per 100K population (numeric - decimal) \\
\end{longtable}
\section{Additional experiment results}
\label{appendix_additional}
In this section, we present additional experimental results, including those for logistic regression, which complement the linear regression results shown in the main paper. This section also provides trace plots and Gelman-Rubin statistic ($\hat{R}$) values for convergence analysis, as well as marginal distribution plots with additional priors included.

\begin{figure}[htb]
\centering
\begin{subfigure}[t]{\linewidth}
    \centering
    \includegraphics[width=0.75\linewidth]{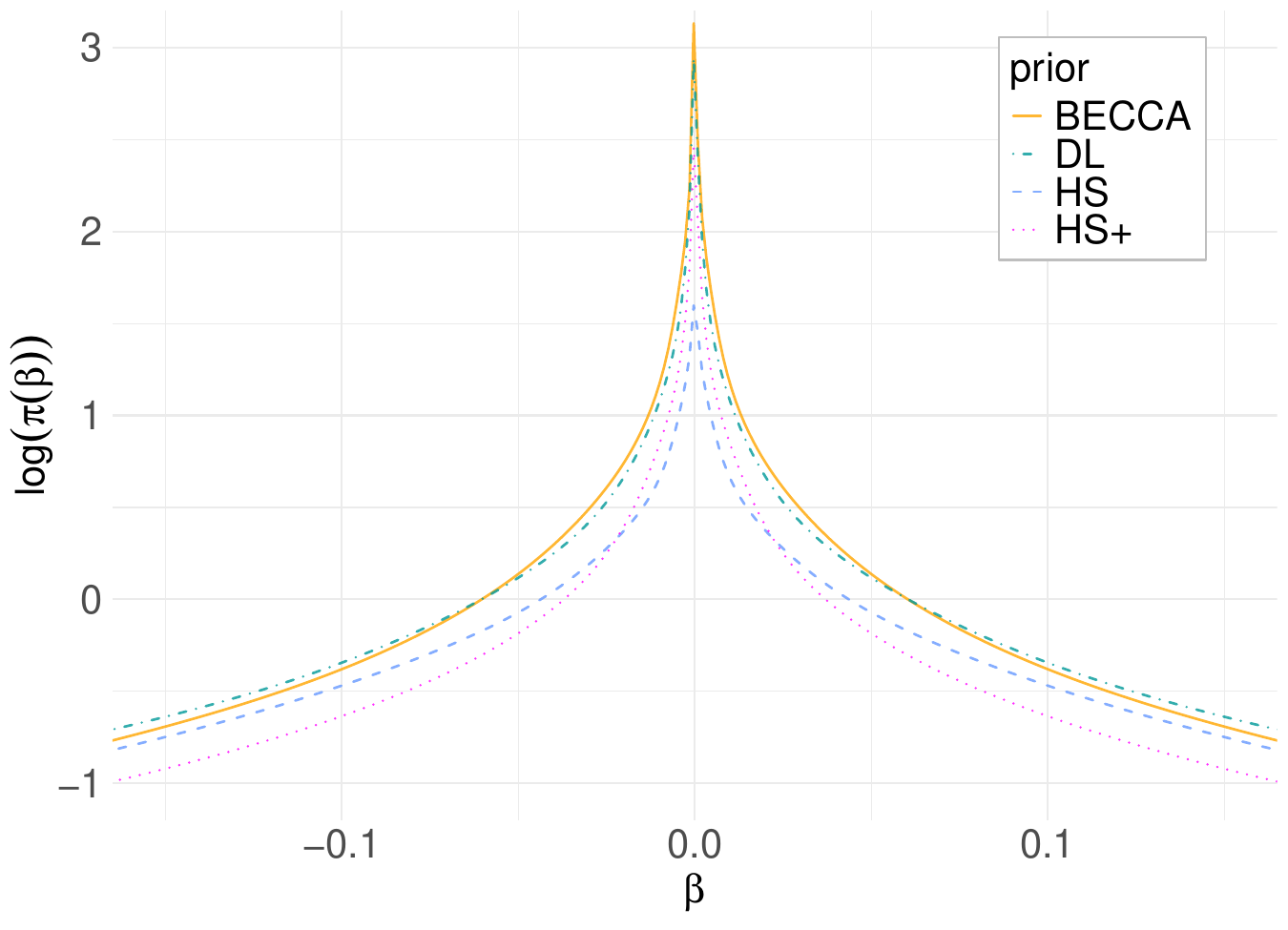}
    \caption{}
    \label{fig:LogMarginalBetaOriginAll}
\end{subfigure}

\vspace{1cm}  

\begin{subfigure}[t]{\linewidth}
    \centering
    \includegraphics[width=0.65\linewidth]{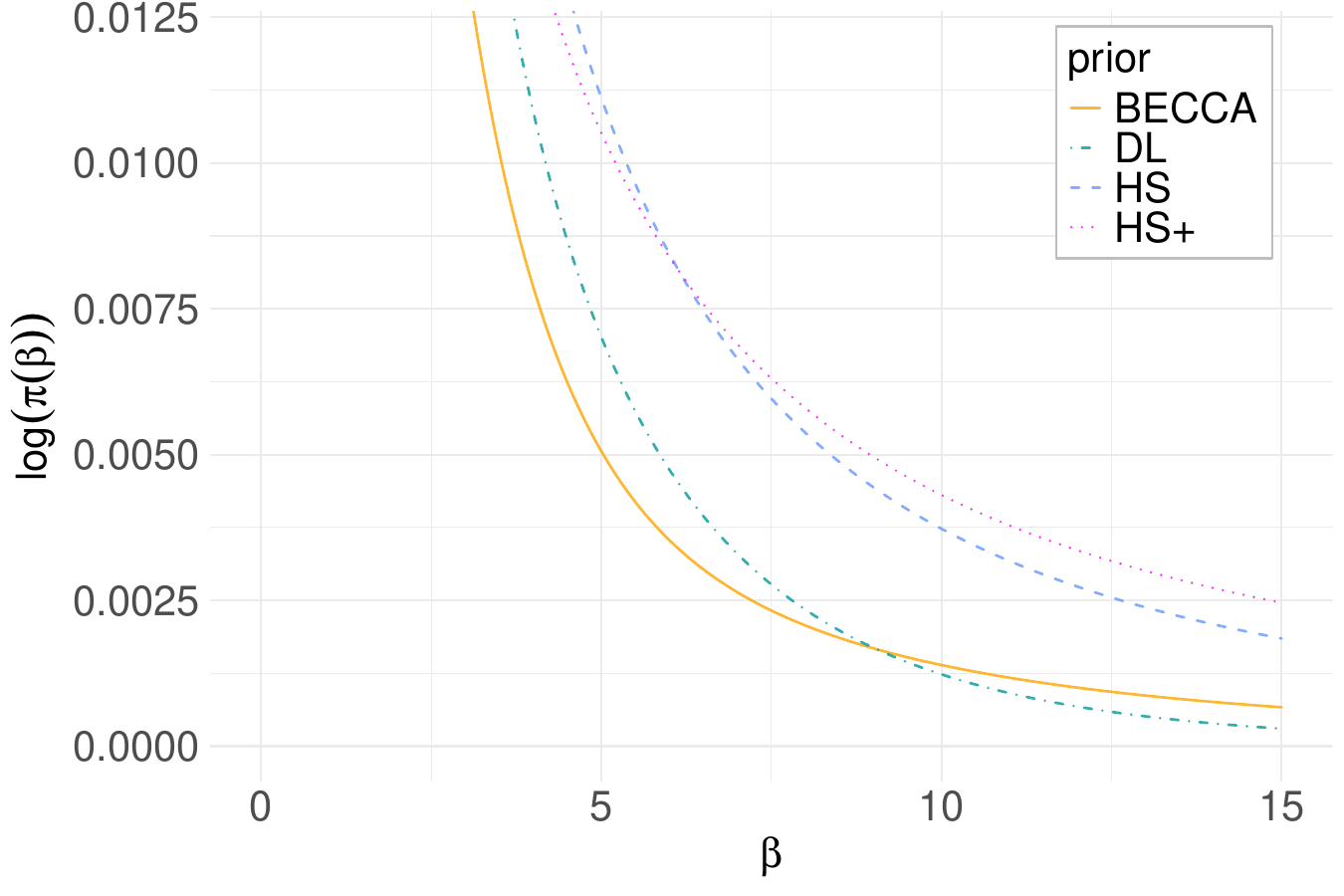}
    \caption{}
    \label{fig:MarginalBeta_tailAll}
\end{subfigure}

\vspace{.2in}
\caption{Marginal prior densities of $\beta_j$ under different priors (BECCA, HS+, HS, and DL): (a) near the origin on a linear scale, (b) near the origin on a log scale and (c) the tail regions on a linear scale.}
\end{figure}

\begin{figure}[htb]
\centering
\begin{subfigure}[t]{0.44\linewidth}
    \centering
    \includegraphics[width=\linewidth]{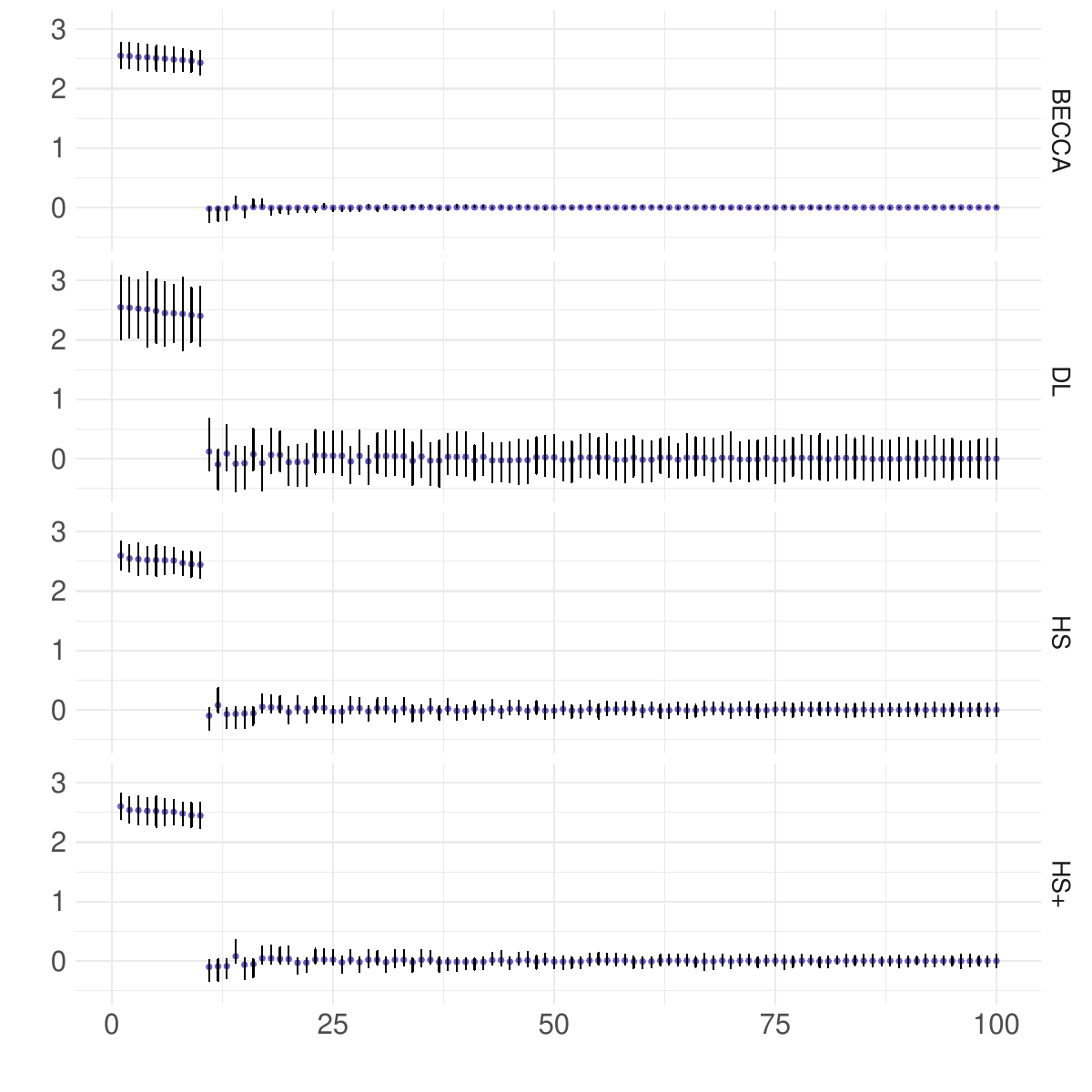}
    \caption{}
    \label{fig:95CI_beta_2.5_rep1}
\end{subfigure}
\hspace{1cm}
\begin{subfigure}[t]{0.44\linewidth}
    \centering
    \includegraphics[width=\linewidth]{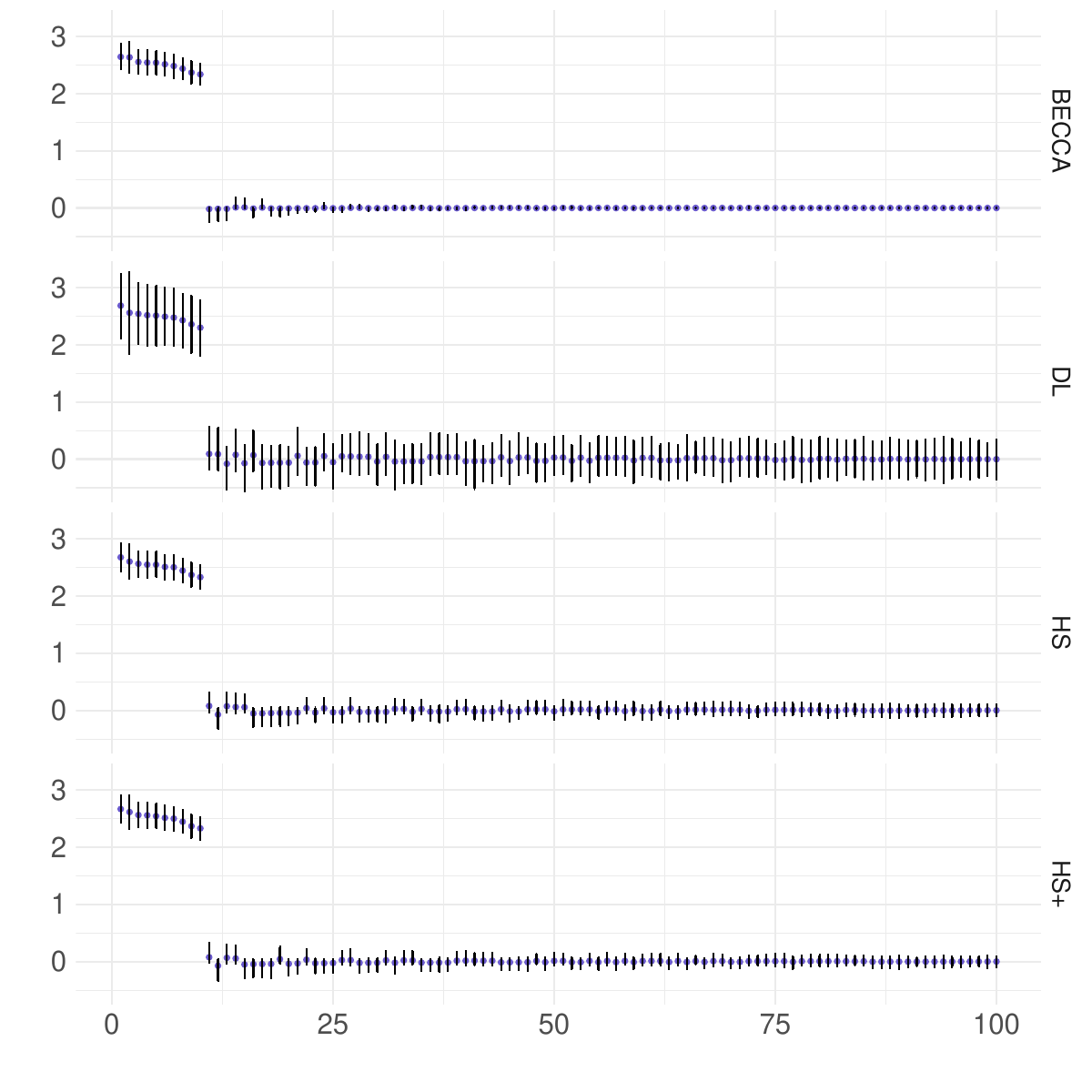}
    \caption{}
    \label{fig:95CI_beta_2.5_rep2}
\end{subfigure}
\caption{Estimated  $\beta_j$ for BECCA, HS, HS+ and DL in linear regression with 100 predictors, where the first 10 true $\beta_j$ are set to 2.5 and the rest to 0. The middle $95\%$ posterior credible intervals are shown as black solid lines, and the posterior means are indicated by blue dots.}
\label{fig:95CI_beta_2.5}
\end{figure}

\begin{figure}[htb]
\centering
\begin{subfigure}[t]{0.44\linewidth}
    \centering
    \includegraphics[width=\linewidth]{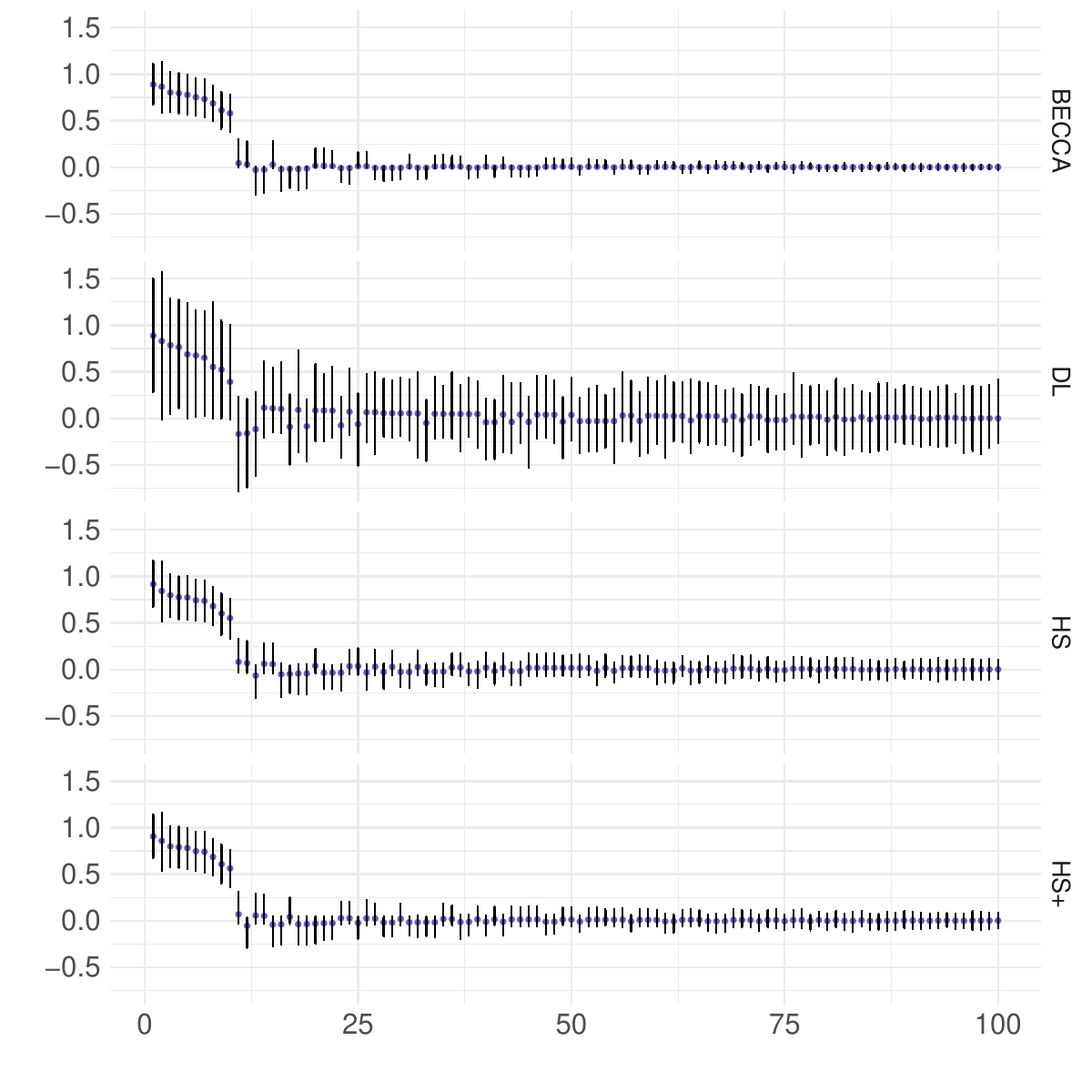}
    \caption{}
    \label{fig:95CI_beta_0.75_rep1}
\end{subfigure}
\hspace{1cm}
\begin{subfigure}[t]{0.44\linewidth}
    \centering
    \includegraphics[width=\linewidth]{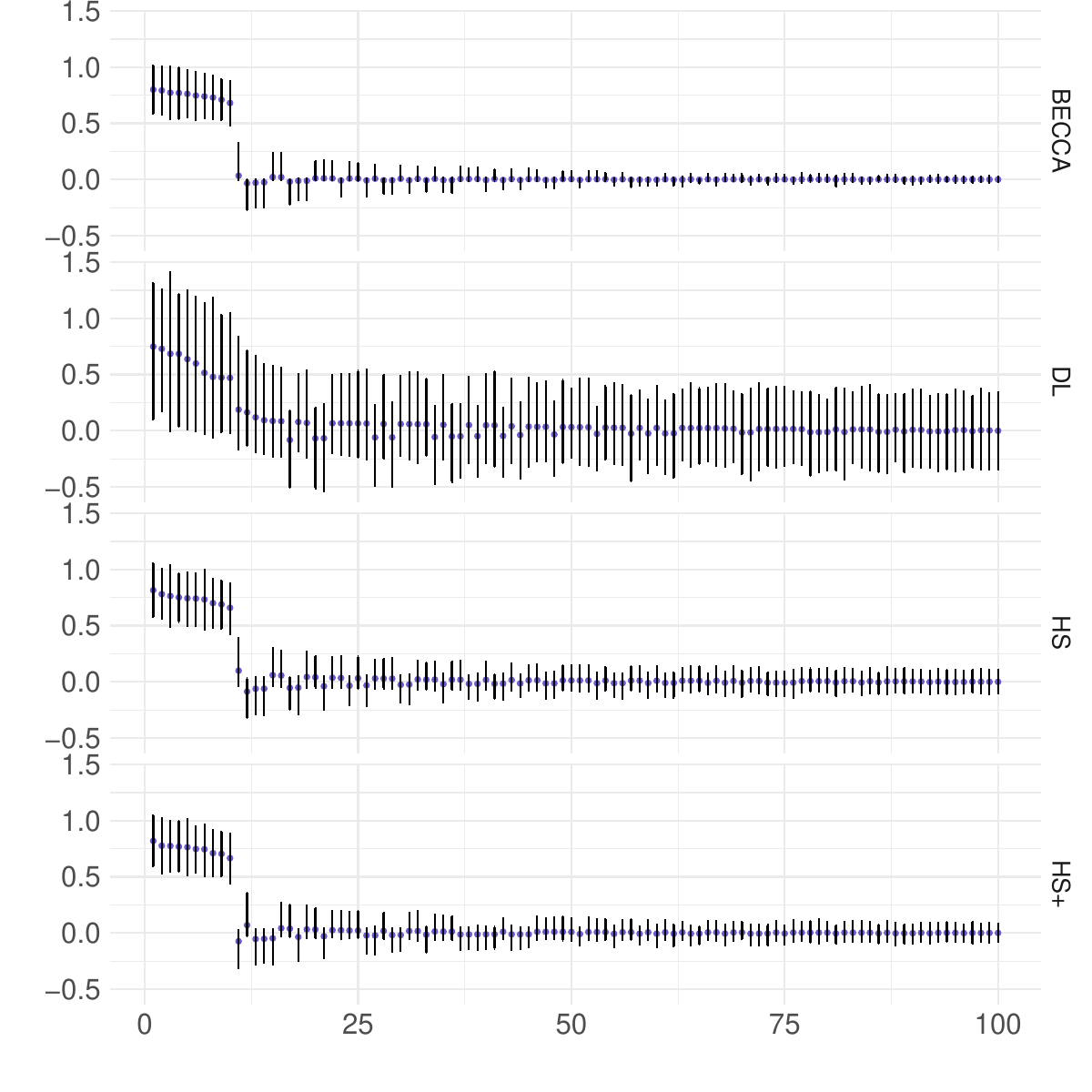}
    \caption{}
    \label{fig:95CI_beta_0.75_rep2}
\end{subfigure}
\caption{Estimated  $\beta_j$ for BECCA, HS, HS+ and DL in linear regression with 100 predictors, where the first 10 true $\beta_j$ are set to 0.75 and the rest to 0. The middle $95\%$ posterior credible intervals are shown as black solid lines, and the posterior means are indicated by blue dots.}
\label{fig:95CI_beta_0.75}
\end{figure}

\begin{figure}[htb]
\centering
\begin{subfigure}[t]{0.44\linewidth}
    \centering
    \includegraphics[width=\linewidth]{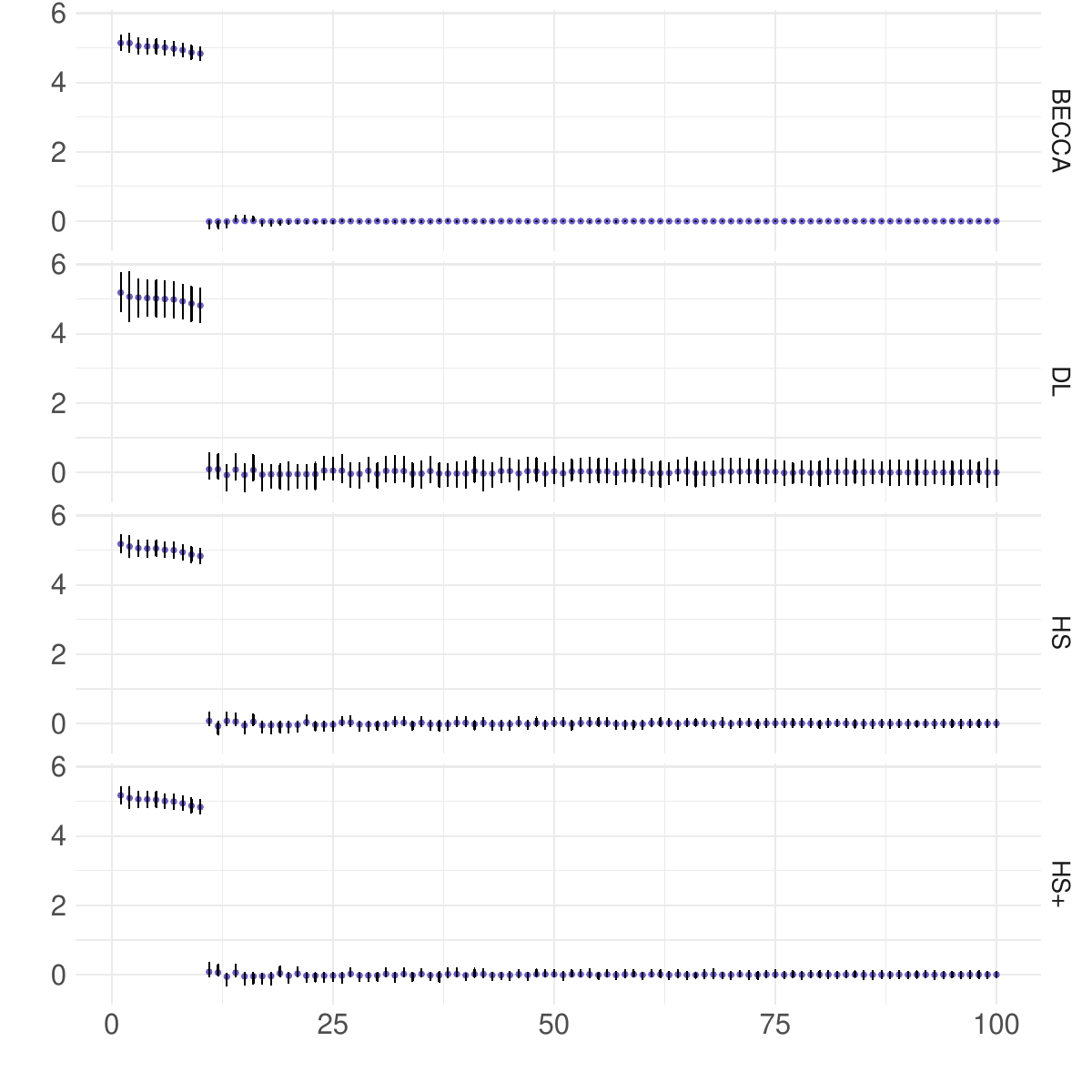}
    \caption{}
    \label{fig:95CI_beta_5_rep1}
\end{subfigure}
\hspace{1cm}
\begin{subfigure}[t]{0.44\linewidth}
    \centering
    \includegraphics[width=\linewidth]{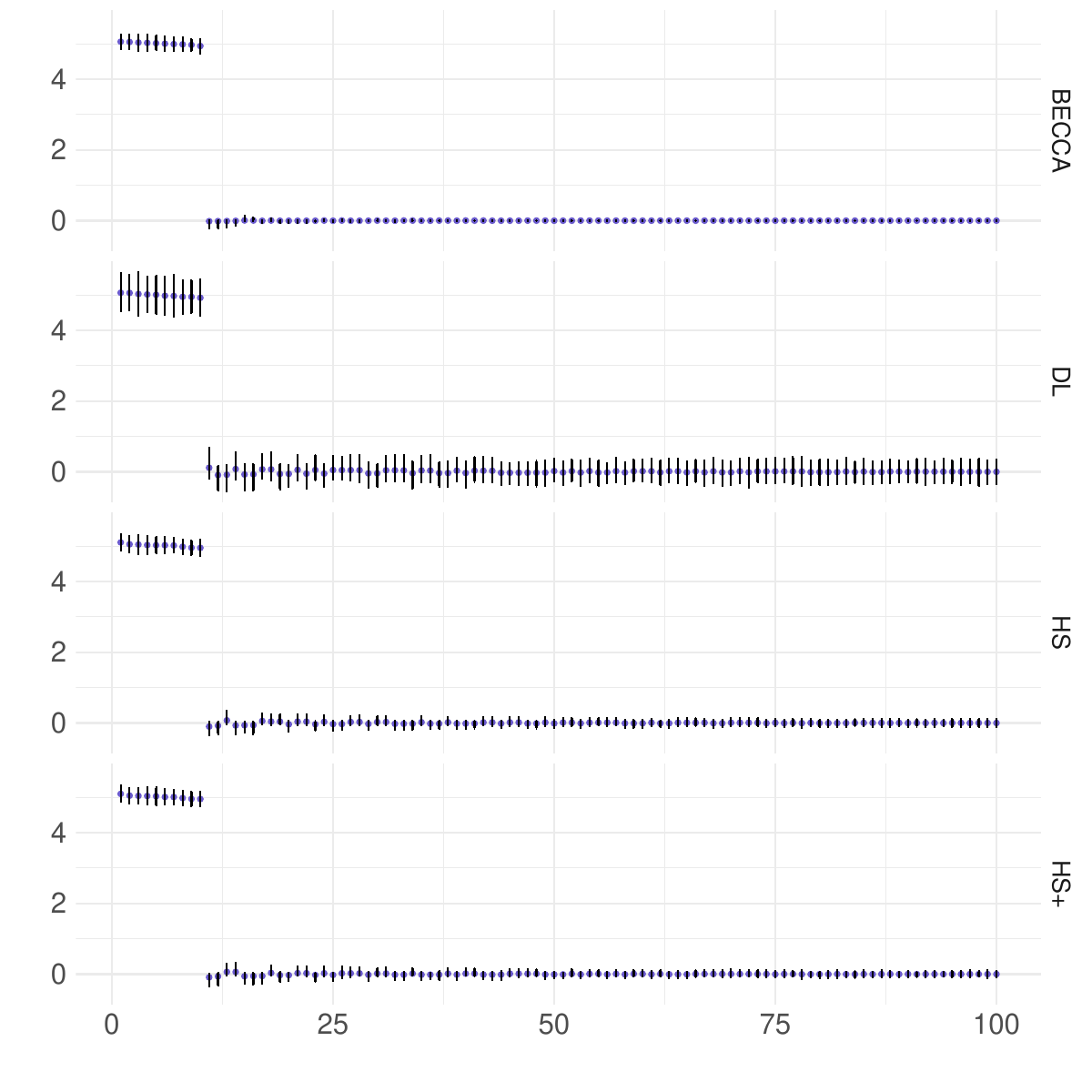}
    \caption{}
    \label{fig:95CI_beta_5_rep2}
\end{subfigure}
\caption{Estimated $\beta_j$  for BECCA, HS, HS+ and DL in linear regression with 100 predictors, where the first 10 true $\beta_j$ are set to 5 and the rest to 0. The middle $95\%$ posterior credible intervals are shown as black solid lines, and the posterior means are indicated by blue dots.}
\label{fig:95CI_beta_5}
\end{figure}

\begin{figure}[htb]
\centering
\begin{subfigure}[t]{0.44\linewidth}
    \centering
    \includegraphics[width=\linewidth]{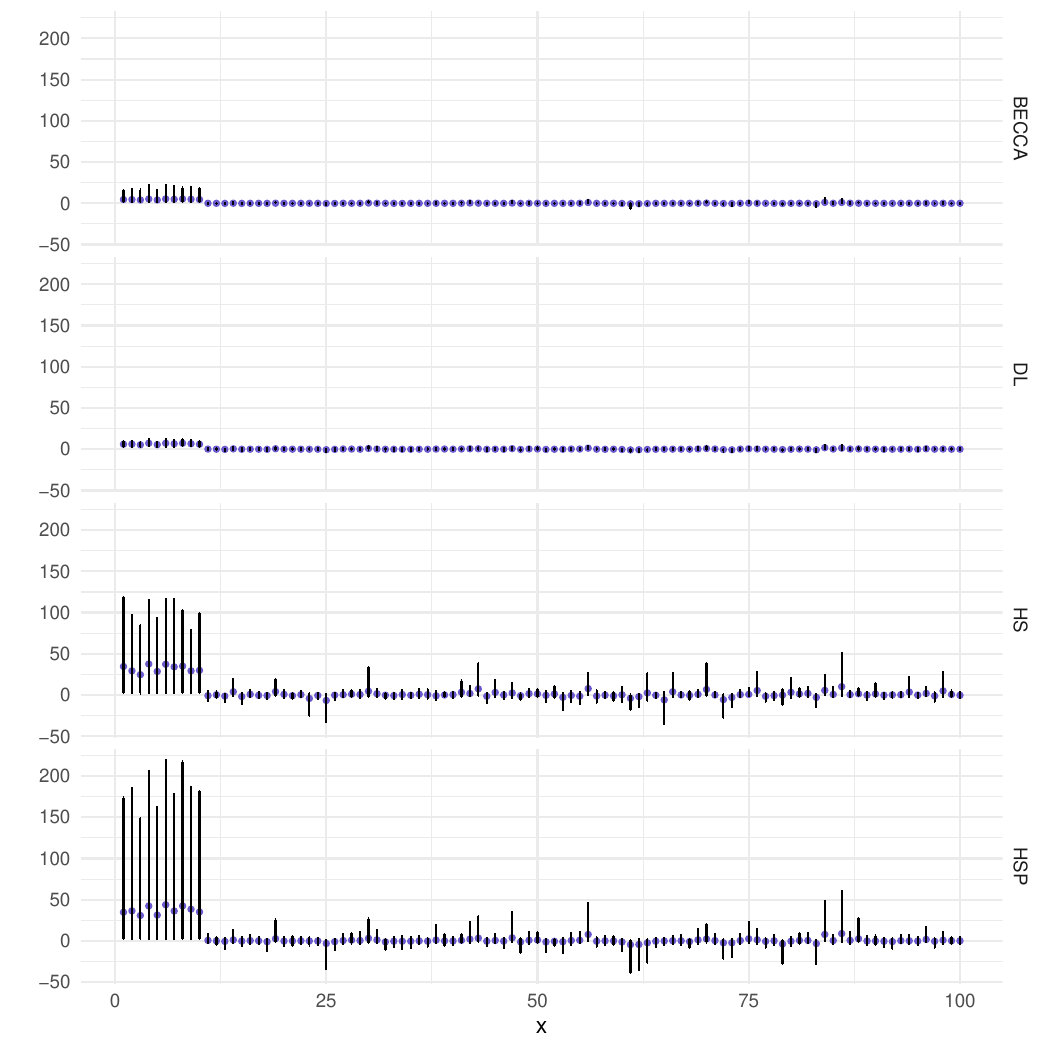}
    \caption{}
    \label{fig:95CI_betaAll_logit}
\end{subfigure}
\hspace{1cm}
\begin{subfigure}[t]{0.44\linewidth}
    \centering
    \includegraphics[width=\linewidth]{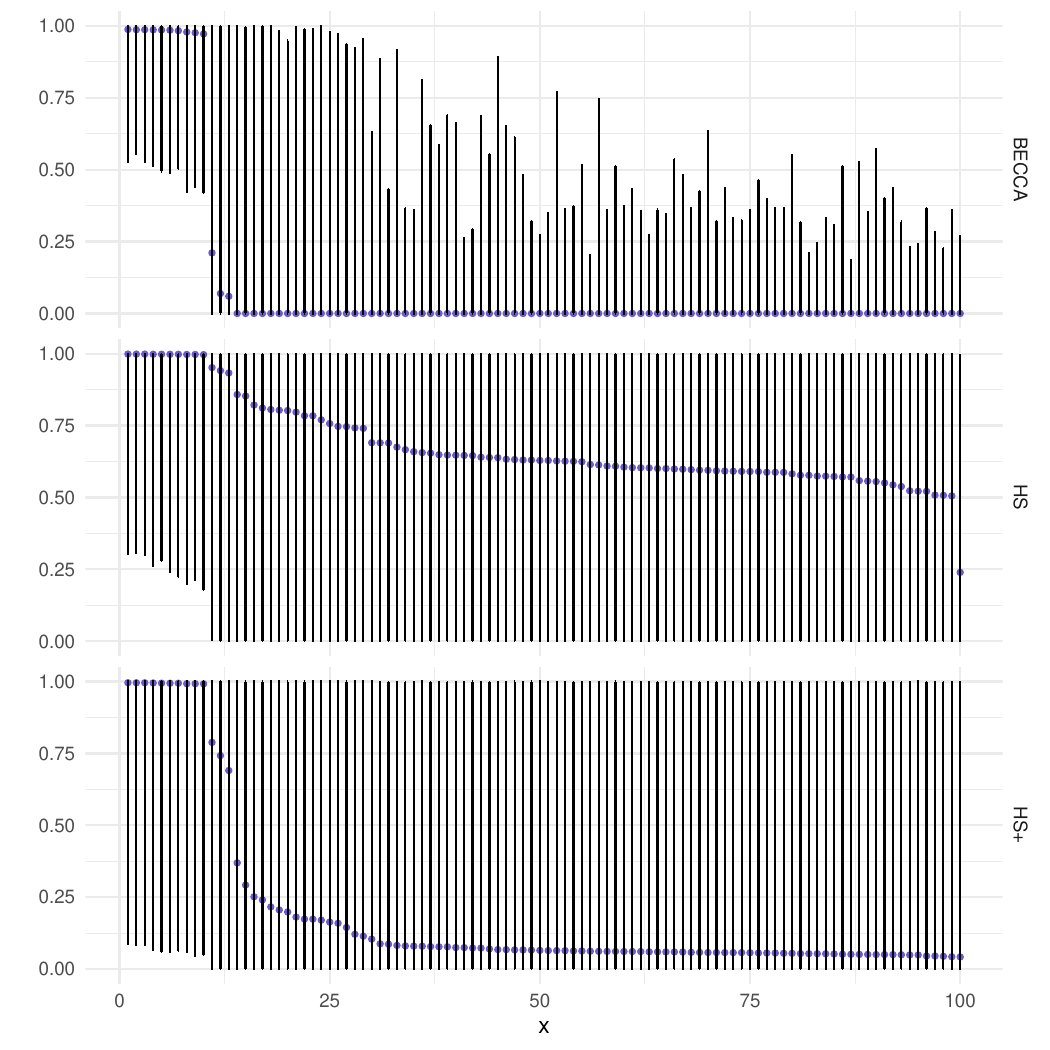}
    \caption{}
    \label{fig:95CI_gamma_logit}
\end{subfigure}
\caption{Estimated (a) $\beta_j$ and (b) $\gamma_j$ for BECCA, HS, HS+ and DL in logistic regression with 100 predictors, where the first 10 true $\beta_j$ are set to 2.5 and the rest to 0. The middle $95\%$ posterior credible intervals are shown as black solid lines, and the posterior means are indicated by blue dots.}
\end{figure}